\documentclass[twocolumn]{aastex61}

\usepackage{xspace}
\usepackage{color}
\usepackage{graphicx}

\newcommand{\nraoblurb}{The National Radio Astronomy Observatory is
a facility of the National Science Foundation operated under cooperative
agreement by Associated Universities, Inc.}
\newcommand{\nexpo}[2]{\ensuremath{{#1}\times10^{#2}}\xspace}

\newcommand{\gsim}{\ensuremath{\,\gtrsim\,}\xspace}

\newcommand{\vlsr}{\ensuremath{V_{\rm LSR}}\xspace}

\newcommand{\kms}{\ensuremath{\,{\rm km\,s^{-1}}}\xspace}

\newcommand{\m}{\ensuremath{\,{\rm m}}\xspace}
\newcommand{\cm}{\ensuremath{\,{\rm cm}}\xspace}
\newcommand{\pc}{\ensuremath{\,{\rm pc}}\xspace}
\newcommand{\kpc}{\ensuremath{\,{\rm kpc}}\xspace}
\newcommand{\K}{\ensuremath{\,{\rm K}}\xspace}

\newcommand{\khz}{\ensuremath{\,{\rm kHz}}\xspace}
\newcommand{\mhz}{\ensuremath{\,{\rm MHz}}\xspace}
\newcommand{\ghz}{\ensuremath{\,{\rm GHz}}\xspace}

\newcommand{\degree}{\ensuremath{\,^\circ}\xspace}
\newcommand{\degper}{\ensuremath{\rlap.{^{\circ}}}}

\newcommand{\msun}{\ensuremath{{\rm \,M_\odot}}\xspace}     
 
\newcommand{\hi}{H\,{\sc i}}
\newcommand{\hii}{H\,{\sc ii}}

\newcommand{\co} {\ensuremath{^{\rm 12}{\rm CO}}\xspace}
\newcommand{\cor}{\ensuremath{^{\rm 13}{\rm CO}}\xspace}

\revised{\today\ Submitted -- TVW}

\shorttitle{CO in the OSC}
\shortauthors{Wenger et al.}

\begin{document}

\title{Carbon Monoxide Observations Toward Star Forming Regions in the
  Outer Scutum-Centaurus Spiral Arm}

\author{Trey V. Wenger}

\affiliation{Astronomy Department, University of Virginia, P.O. Box
  400325, Charlottesville, VA 22904-4325, USA.}

\affiliation{National Radio Astronomy Observatory, 520 Edgemont Rd.,
  Charlottesville, VA 22903, USA.}

\author{Asad A. Khan}

\affiliation{Astronomy Department, University of Virginia, P.O. Box
  400325, Charlottesville, VA 22904-4325, USA.}

\author{Nicholas G. Ferraro}

\affiliation{Astronomy Department, University of Virginia, P.O. Box
  400325, Charlottesville, VA 22904-4325, USA.}

\author{Dana S. Balser}

\affiliation{National Radio Astronomy Observatory, 520 Edgemont Rd.,
  Charlottesville, VA 22903, USA.}

\author{W. P. Armentrout}

\affiliation{Department of Physics and Astronomy, West Virginia
  University, PO Box 6315, Morgantown WV 26506, USA.}

\affiliation{Center for Gravitational Waves and Cosmology, West
  Virginia University, Chestnut Ridge Research Building, Morgantown,
  WV 26505, USA.}

\author{L. D. Anderson}

\affiliation{Department of Physics and Astronomy, West Virginia
  University, PO Box 6315, Morgantown WV 26506, USA.}

\affiliation{Adjunct Astronomer at the Green Bank Observatory,
  P.O. Box 2, Green Bank WV 24944, USA.}

\affiliation{Center for Gravitational Waves and Cosmology, West
  Virginia University, Chestnut Ridge Research Building, Morgantown,
  WV 26505, USA.}

\author{T. M. Bania}

\affiliation{Institute for Astrophysical Research, Department of
  Astronomy, Boston University, 725 Commonwealth Avenue, Boston, MA,
  02215, USA.}

\begin{abstract}
  The Outer Scutum-Centaurus arm (OSC) is the most distant molecular
  spiral arm known in the Milky Way. The OSC may be the very distant
  end of the well-known Scutum-Centaurus arm, which stretches from the
  end of the Galactic bar to the outer Galaxy.  At this distance the
  OSC is seen in the first Galactic quadrant. The population of star
  formation tracers in the OSC remains largely uncharacterized.
  Extragalactic studies show a strong correlation between molecular
  gas and star formation, and carbon monoxide (CO) emission was
  recently discovered in the OSC.  Here we use the Arizona Radio
  Observatory (ARO) 12\m\ telescope to observe the \co\ J = 1--0 and
  \cor\ J = 1--0 transitions toward 78 \hii\ region candidates chosen
  from the {\it WISE} Catalog of Galactic \hii\ Regions. These targets
  are spatially coincident with the Galactic longitude-latitude
  ($\ell, b$) OSC locus as defined by \hi\ emission. We detect CO
  emission in $\sim 80$\% of our targets.  In total, we detect 117 \co
  and 40 \cor emission lines. About 2/3 of our targets have at least
  one emission line originating beyond the Solar orbit. Most of the
  detections beyond the Solar orbit are associated with the Outer Arm,
  but there are 17 \co\ emission lines and 8 \cor\ emission lines with
  LSR velocities that are consistent with the velocities of the
  OSC. There is no apparent difference between the physical properties
  (e.g., molecular column density) of these OSC molecular clouds and
  non--OSC molecular clouds within our sample.
\end{abstract}


\section{Introduction}

Galactic structure in the Milky Way is difficult to determine because
of our location within the disk and the difficulties in deriving
accurate distances.  There is strong evidence, however, that the Milky
Way is a barred spiral galaxy \citep[e.g.,][]{churchwell2009}.  Most
barred spiral galaxies are Grand Design galaxies, with two prominent
symmetric spiral arms \citep{elmegreen1982}.  Understanding the Milky
Way structure is important since bars and spiral arms have dynamical
effects (e.g., radial migration) and influence star formation (e.g.,
shock gas).

To our knowledge, \citet{dame2011} were the first to connect the
Scutum-Centaurus (SC) spiral arm from its beginning at the end of the
bar in the inner Galaxy to the outer Galaxy in the first quadrant.
They called this outermost section the Outer Scutum-Centaurus (OSC)
arm. In the first quadrant, the OSC is at a distance between
\(15\kpc\) and \(19\kpc\) kpc from the Galactic Center (GC) and
between \(20\kpc\) and \(25\kpc\) from the Sun \citep{armentrout2017}.
The connection between the SC and OSC arm segments, however, remains
hypothetical. Toward the GC, velocity crowding and complex structure
prohibits clear identification of spiral structure.  The \hi\ emission
that has now been associated with the OSC arm was detected long
ago. The arm appears in the early 21 cm maps of both \citet{kerr1969}
and \citet{weaver1974}.  Yet it took decades to trace the SC spiral
arm from the bar to the outer Galaxy.

\citet{dame2011} carefully traced the OSC in the Leiden/Argentine/Bonn
(LAB) \hi\ 21\cm\ line all-sky survey data \citep{hartmann1997,
  arnal2000, bajaja2005, kalberla2005} and then, using the Center for
Astrophysics 1.2 m telescope, detected molecular gas in the arm for
the first time, at 10 locations coincident with HI emission peaks. A
CO map was made of one location revealing a molecular cloud with mass
and radius of \nexpo{5}{4}\msun\ and 47\pc, respectively.
\citet{koo2017} recently re-analyzed the LAB survey data using a
peak-finding algorithm and identified the OSC arm as a 20\kpc\ long
\hi\ structure coincident with several \hii\ regions and molecular
clouds.

\citet{sun2015} suggested a possible extension of the OSC from the
first Galactic quadrant into the second quadrant.  Their results are
part of the Milky Way Imaging Scroll Painting (MWISP) project to map
\co\ and \cor\ from Galactic longitudes $-10^{\degree} < \ell <
250^{\degree}$ and Galactic latitudes $-5^{\degree} < b < 5^{\degree}$
with the Purple Mountain Observatory Delingha 13.7\m\ telescope.  They
detected 72 molecular clouds with masses between
$10^{2}-10^{4}$\msun\ that lie between $120^{\degree} < \ell <
150^{\degree}$ and are roughly consistent with an extension of the OSC
assuming a log-periodic spiral arm model.  First quadrant data from
the MWISP for the longitude range of $35^{\degree} < \ell <
45^{\degree}$ reveal 168 molecular clouds consistent with the OSC
($\ell, v$) locus, defined by $V_{\rm LSR} = -1.6\, \ell \pm
13.2$\kms. These molecular clouds have typical masses and
sizes of \nexpo{3}{3}\msun\ and 5\pc, respectively \citep{sun2017}.

Detection of CO does not necessarily imply star
formation. Observations of high-mass star formation tracers are
therefore critical for characterizing the OSC. \hii\ regions are the
archetypical tracers of star formation and spiral structure. As part
of the \hii\ Region Discovery Survey, \citet{anderson2015} discovered
six \hii\ regions whose radio recombination line (RRL) emission
velocities are consistent with the OSC ($\ell, v$) locus defined by
\(V_{\rm LSR} = -1.6\, \ell \pm 15\kms\).  There are four additional
\hii\ regions in the {\it WISE} Catalog of Galactic \hii\ Regions
consistent with this definition of the OSC \citep{anderson2012}.

\begin{figure*}[!htb]
\centering
\includegraphics[width=0.49\linewidth]{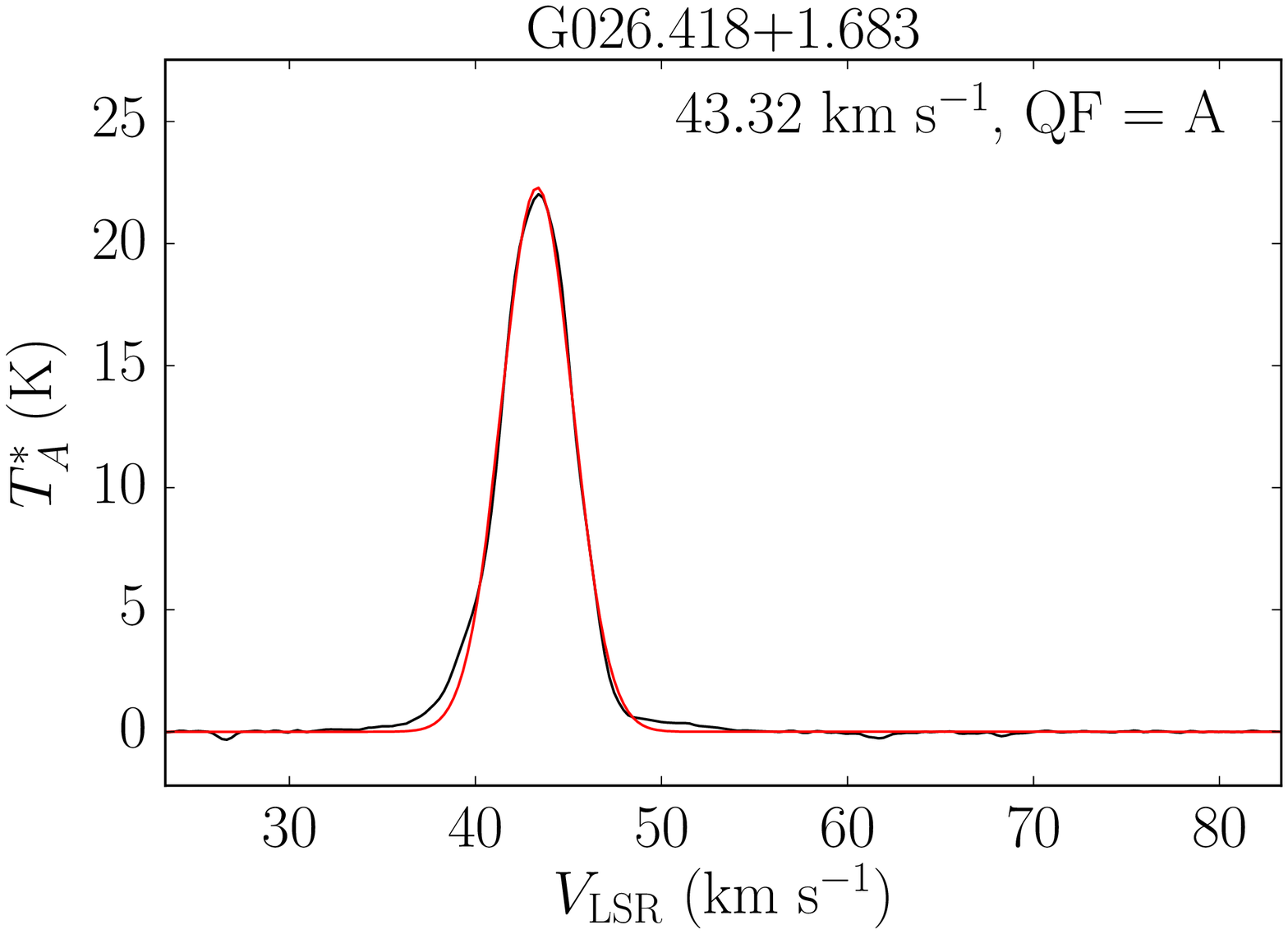}
\includegraphics[width=0.49\linewidth]{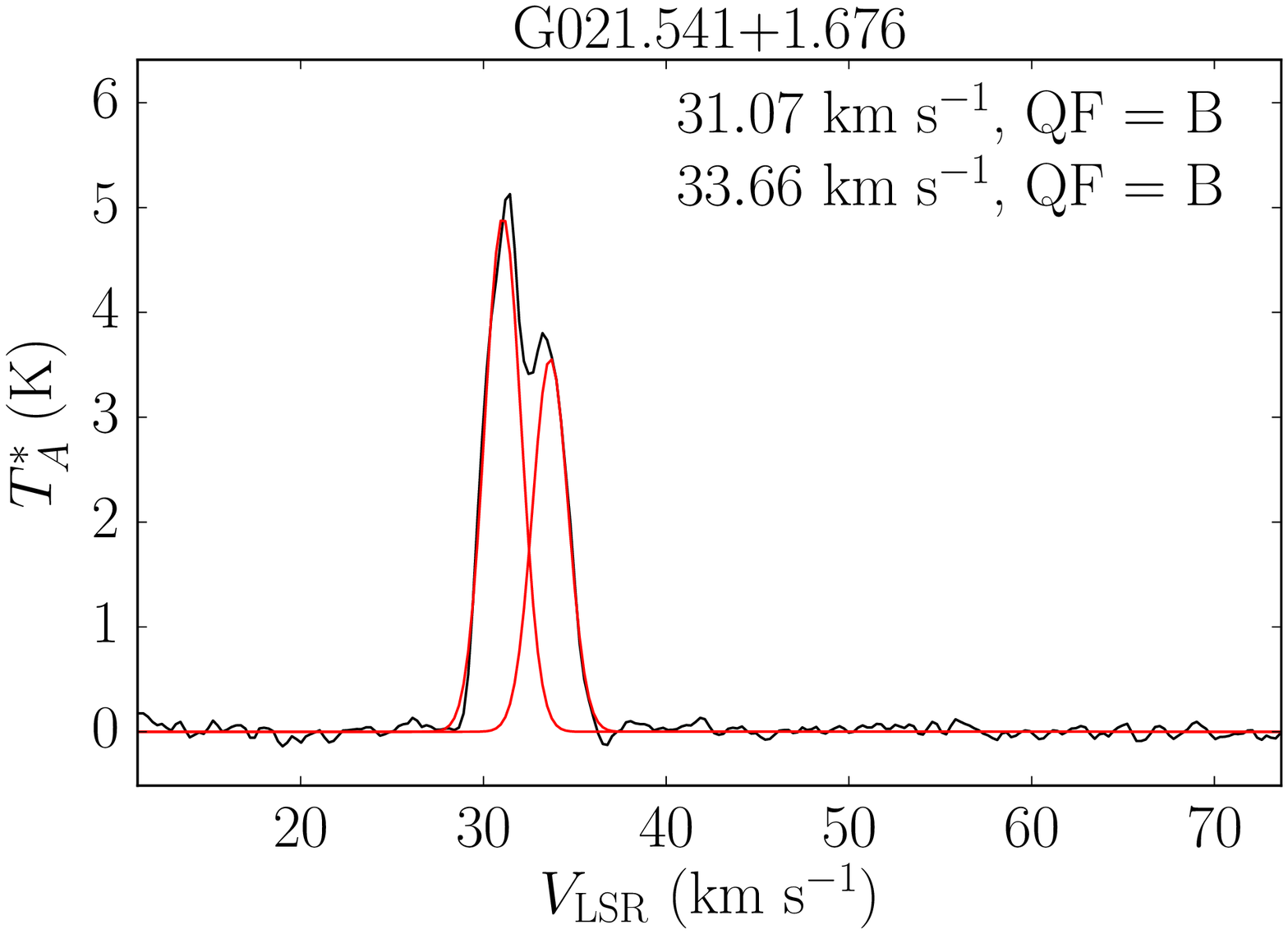}
\includegraphics[width=0.49\linewidth]{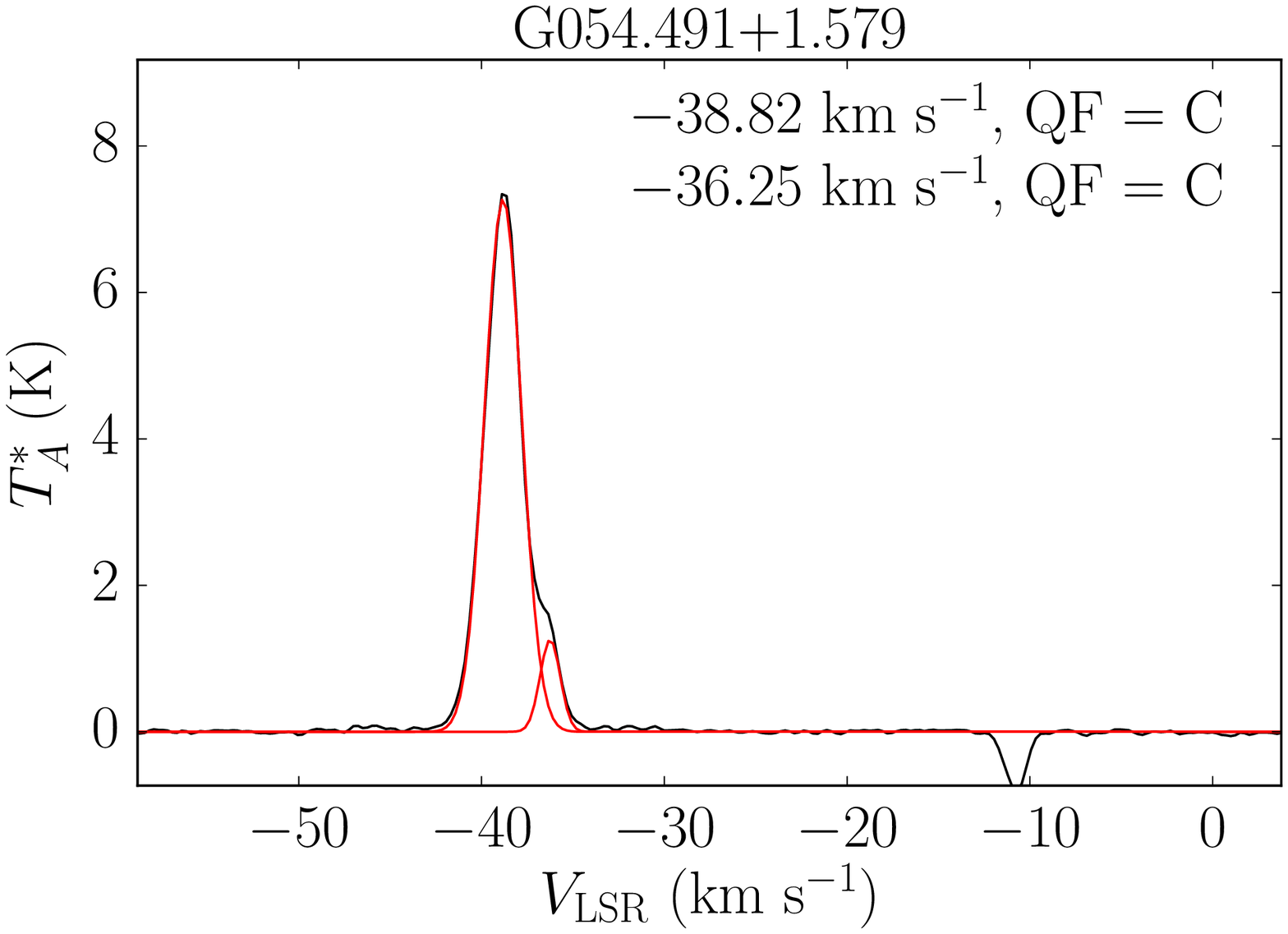}
\includegraphics[width=0.49\linewidth]{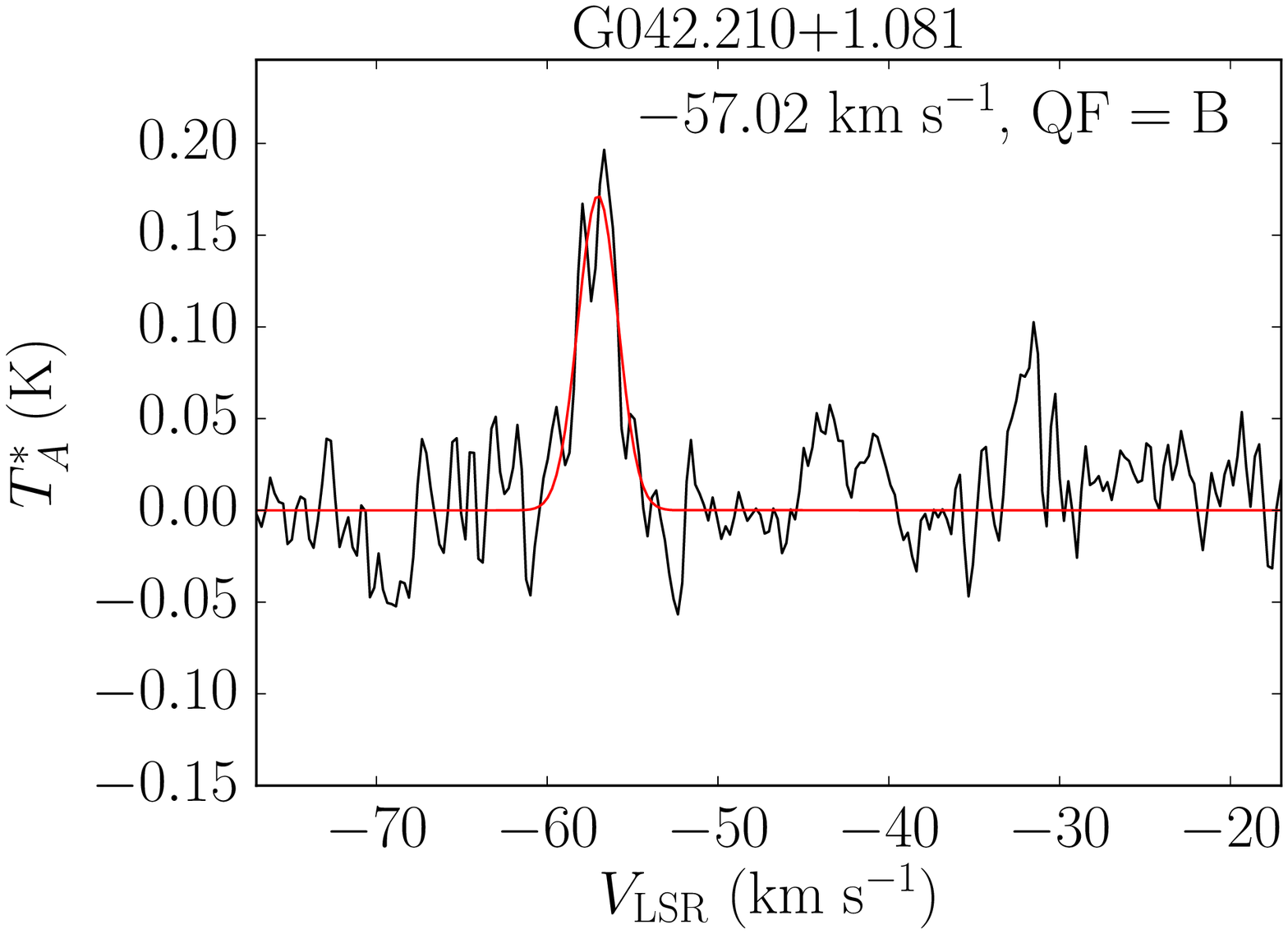}
\caption{Representative \co\ spectra for different quality factors
  (QF).  Plotted is the antenna temperature as a function of the LSR
  velocity.  The black curves are the data and the red curves are
  Gaussian fits to the data.  The LSR velocity of each Gaussian
  profile, together with the QF, is shown in the right-hand corner of the
  plot.}
\label{fig:qf}
\end{figure*}

\begin{figure*}[!htb]
\centering
\includegraphics[width=0.45\linewidth]{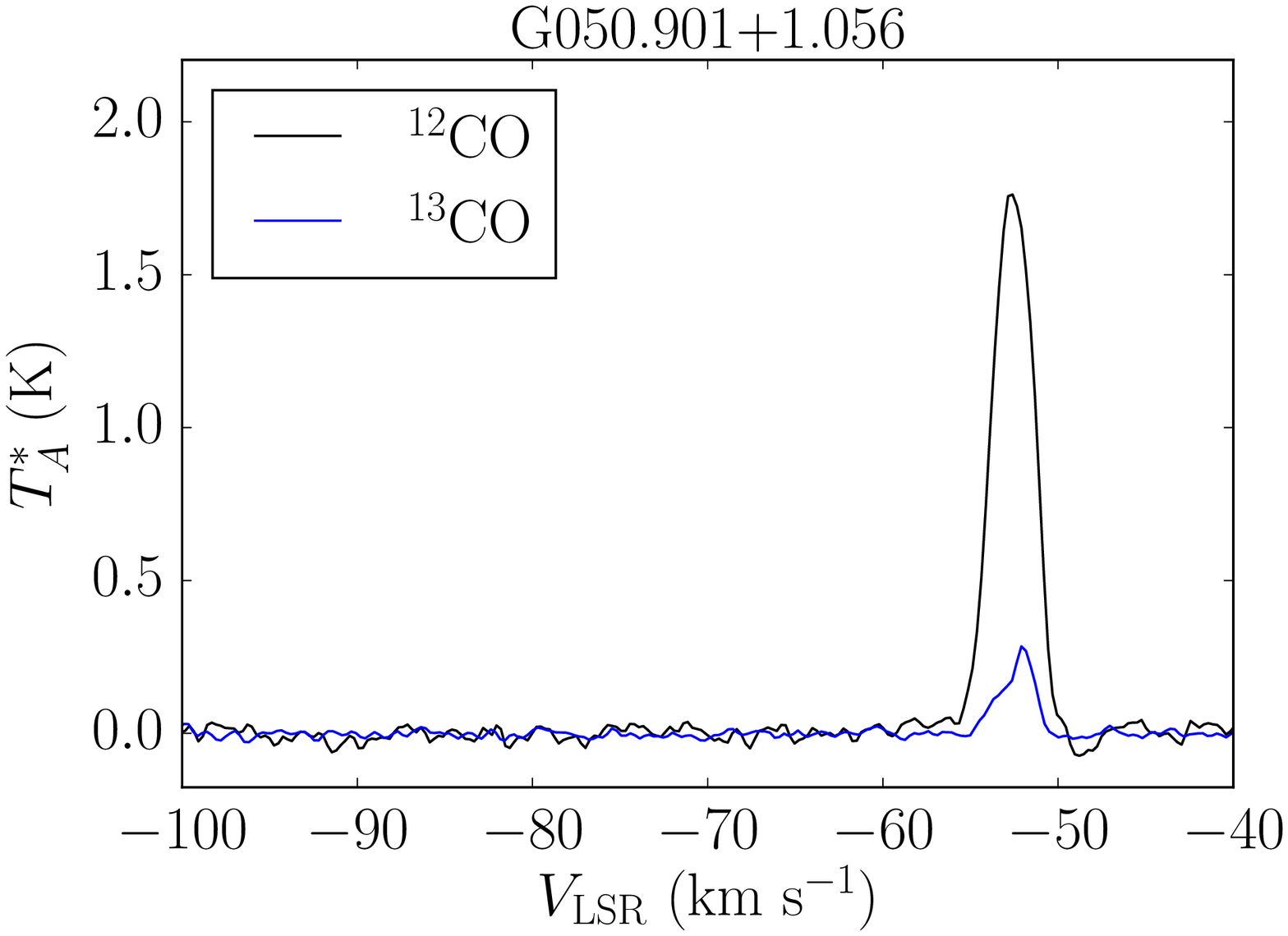}
\includegraphics[width=0.45\linewidth]{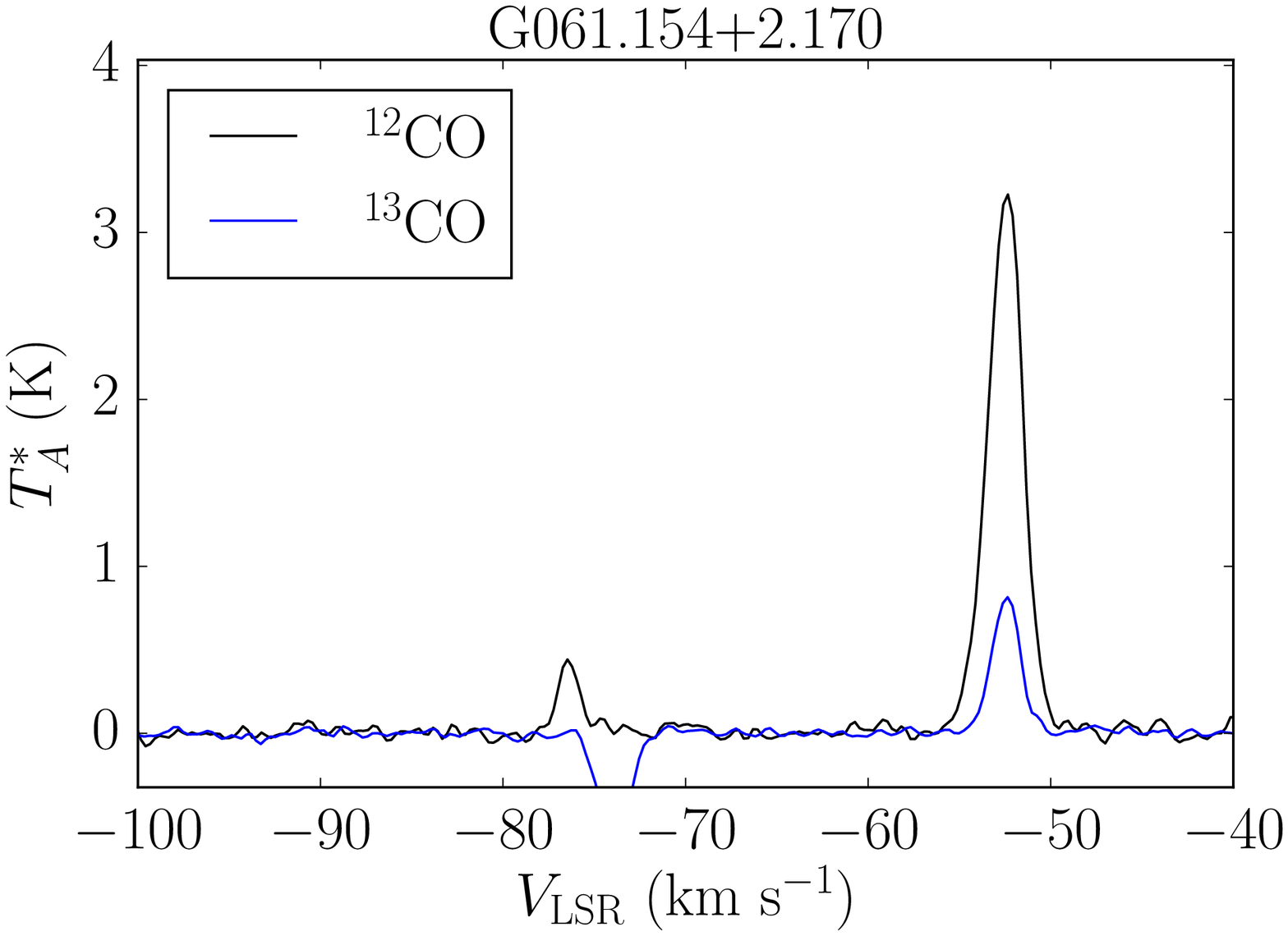} \\
\includegraphics[width=0.45\linewidth]{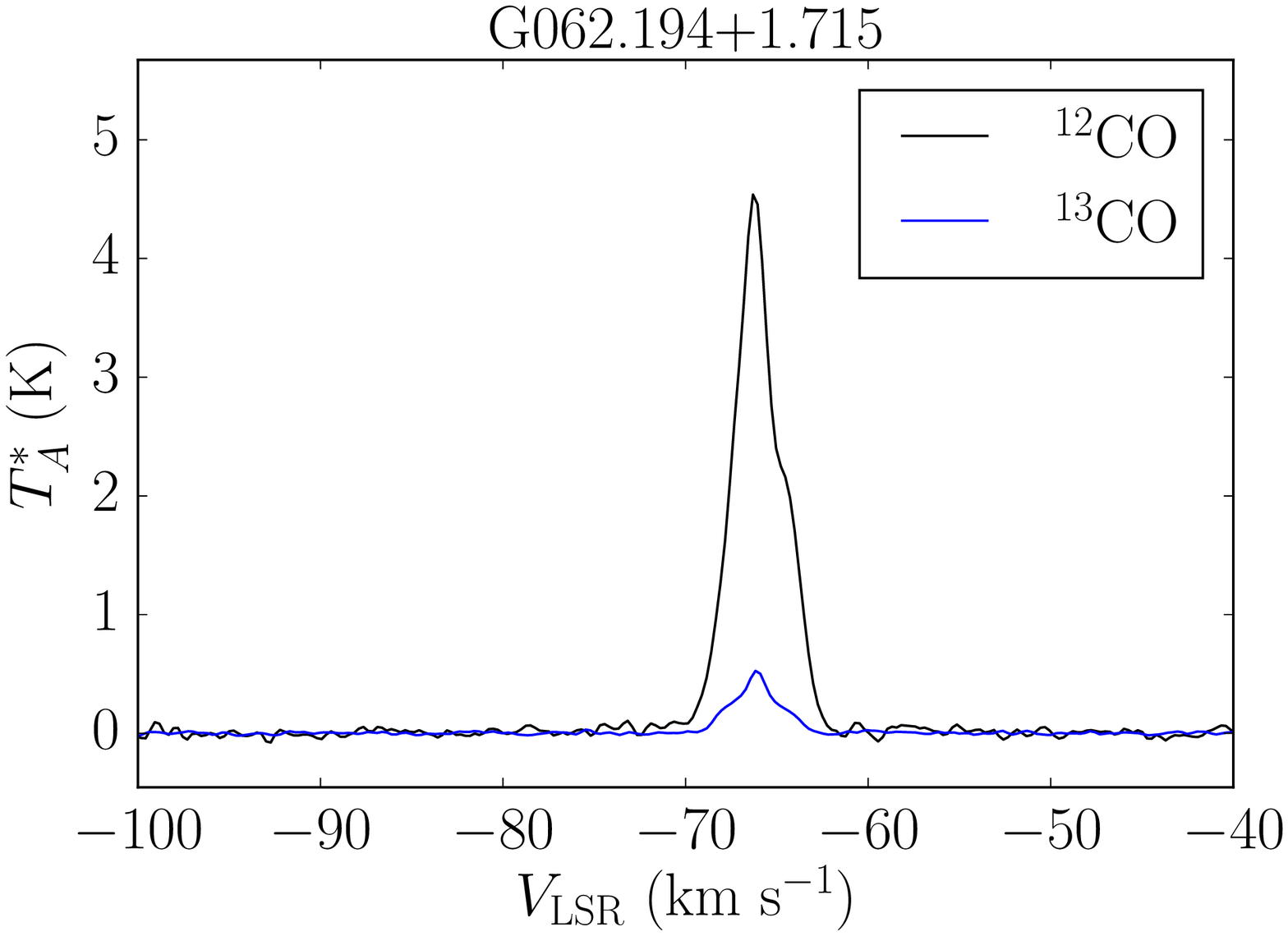}
\includegraphics[width=0.45\linewidth]{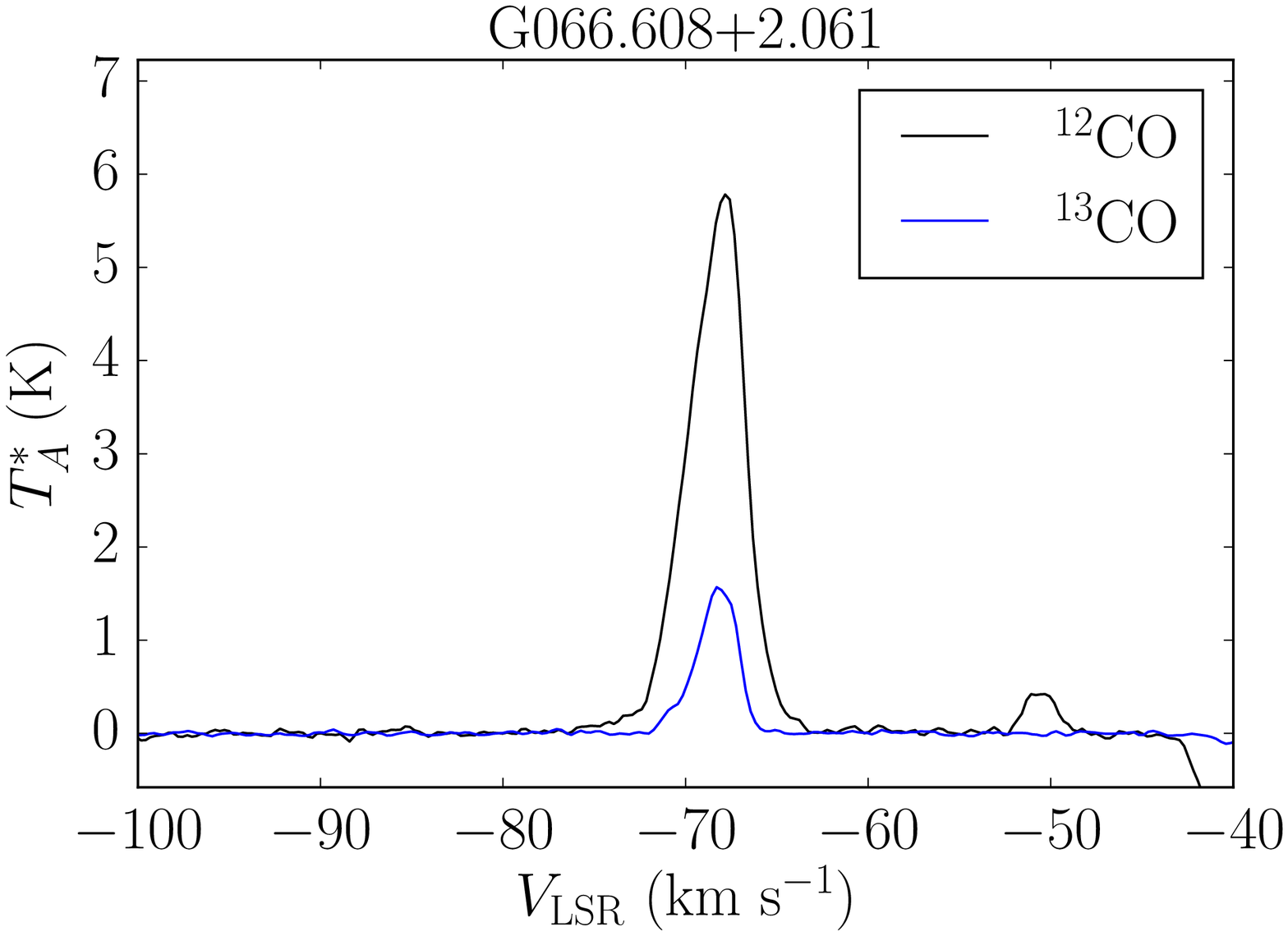}
\caption{\co\ (black) and \cor\ (blue) spectra of targets where the
  association of \co\ emission and \cor\ emission was not
  straightforward.  In G050.901+1.056 and G062.194+1.715 there are two
  \cor\ components that correspond to one \co\ component. In
  G066.608+2.061 there is one \cor\ component that corresponds to two
  \co\ components. Finally, in G061.154+2.170 there are
  \co\ components at velocities where the \cor\ emission is corrupted
  by emission in the Off position.}
\label{fig:special}
\end{figure*}

RRL emission is the best tracer of high-mass star formation, but it is
faint at the large distances of the OSC. \citet{armentrout2017} took a
different approach and used the Green Bank Telescope (GBT) to observe
the dense molecular gas tracers NH$_{3}$ (J,K) = (1,1), (2,2), (3,3)
and H$_{2}$O 6(1,6) $\rightarrow$ 5(2,3) toward 75 {\it WISE}
\hii\ regions and \hii\ region candidates located within the OSC
($\ell, b$) locus as defined by \hi\ \citep{dame2011}. Because these
targets were identified as \hii\ region candidates based on their
mid-infrared morphology, these molecular lines trace dense gas that is
more likely to be associated with star formation. Any detected
spectral line {\it probably} indicates an active star formation
region.  About 20\% of the targets were detected in either ammonia or
water maser emission, but only two have velocities consistent with the
OSC.

\citet{armentrout2017} also observed a similar sample of OSC targets
in radio continuum at 8--10\ghz\ with the Jansky Very Large Array
(JVLA). About 60\% of the targets were detected in radio continuum
with the JVLA. Five of these are associated with the locus of the OSC.
Together, RRL and continuum emission allowed various \hii\ region
physical properties to be derived. Associating the radio continuum
with the molecular transition, however, is less secure since the
molecular cloud may not be associated with the \hii\ region
\citep{anderson2009}. Nonetheless, assuming they are related yields a
distance. If the hydrogen-ionizing photon flux is produced by a single
star, then the observed radio continuum fluxes imply spectral types
ranging from O4 to O8.5. This suggests that even at such large
Galactocentric distances ($> 15$\kpc) high-mass star formation is
ongoing in the Milky Way.

Here we seek to explore the molecular content in the OSC using
\co\ and \cor\ transitions, which are brighter than the molecular line
transitions observed by \citet{armentrout2017}. Since molecular clouds
need not be forming high-mass stars, carbon monoxide is not as good a
tracer of high-mass star formation as RRLs, NH$_{3}$, or H$_{2}$O, but
detection is much more likely at these large distances.  Following
\citet{armentrout2017} we only target \hii\ regions and \hii\ region
candidates from the {\it WISE} Catalog of Galactic
\hii\ Regions. These targets all have the same characteristic infrared
morphology which increases the probability that the molecular gas is
associated with high-mass star formation.

\begin{figure*}[!htb]
\centering
\includegraphics[width=0.32\linewidth]{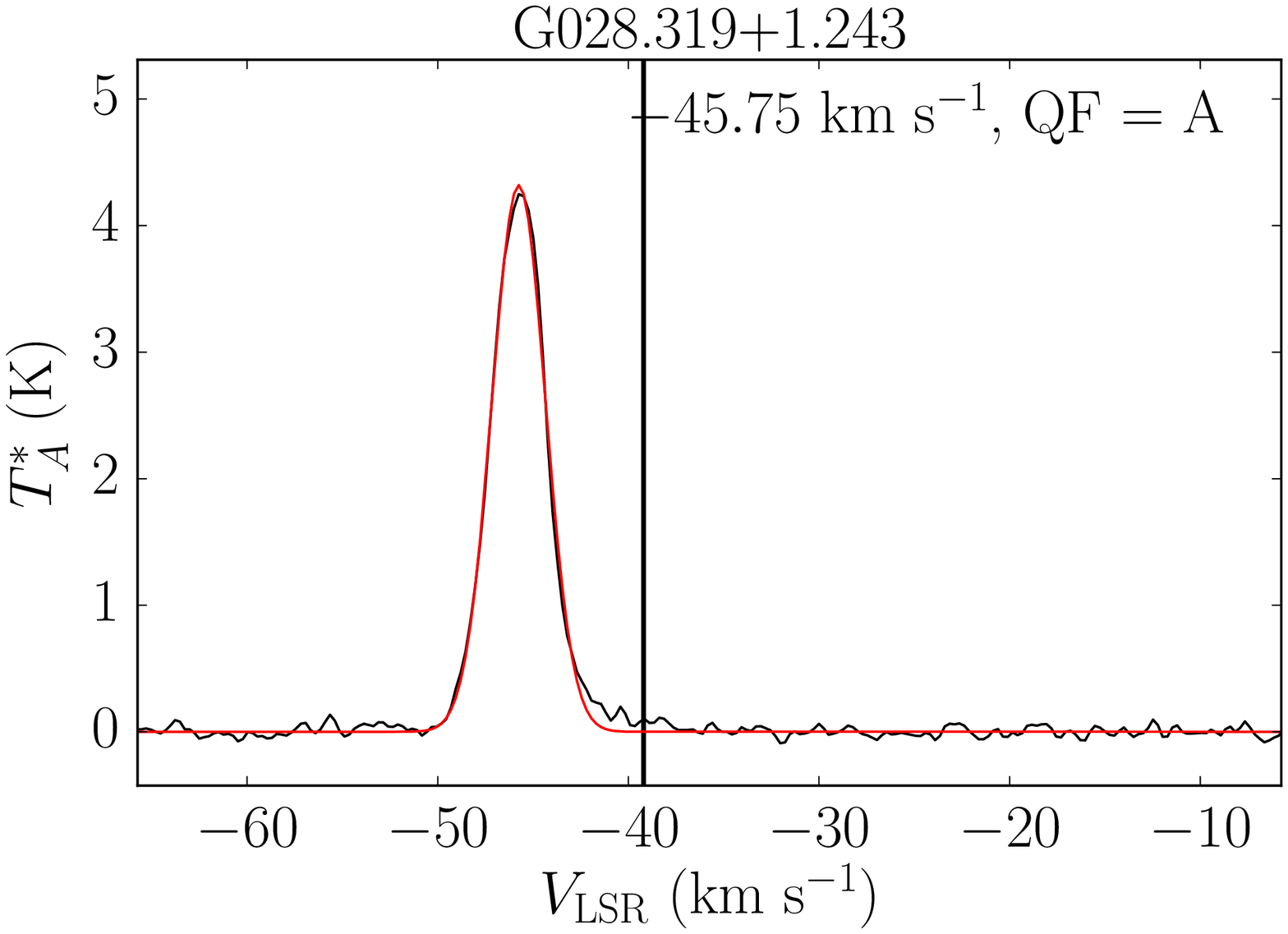}
\includegraphics[width=0.32\linewidth]{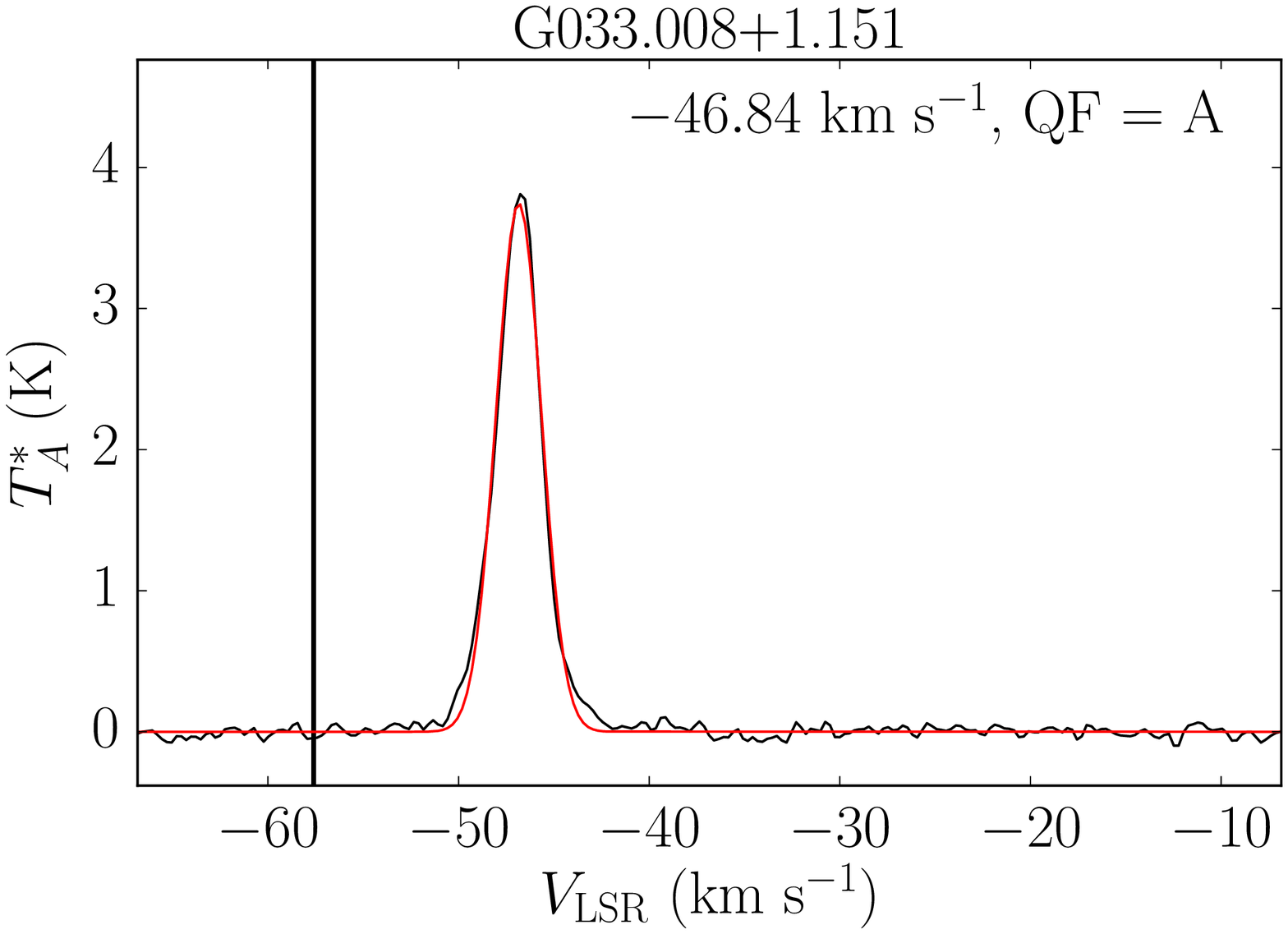}
\includegraphics[width=0.32\linewidth]{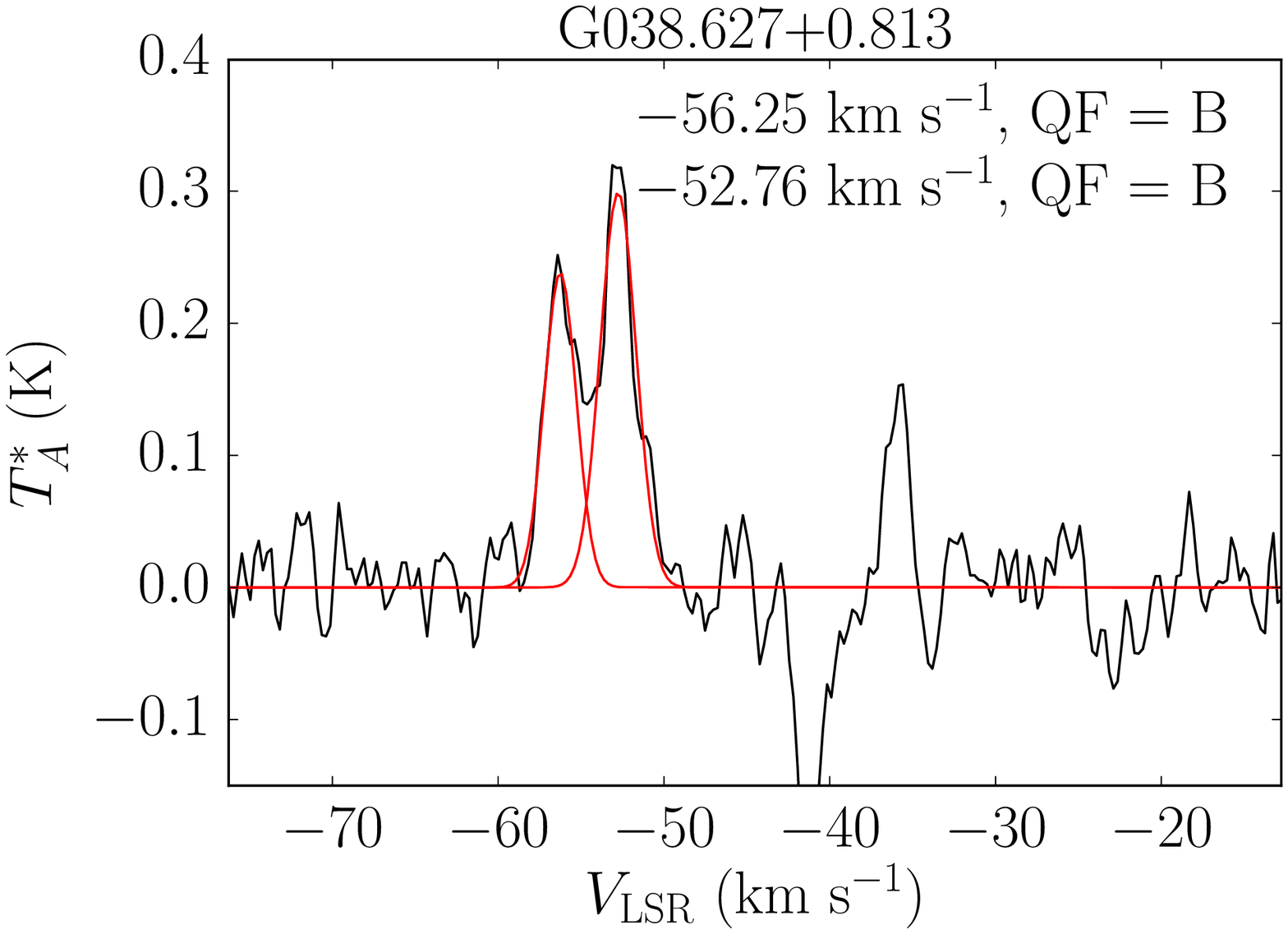}
\includegraphics[width=0.32\linewidth]{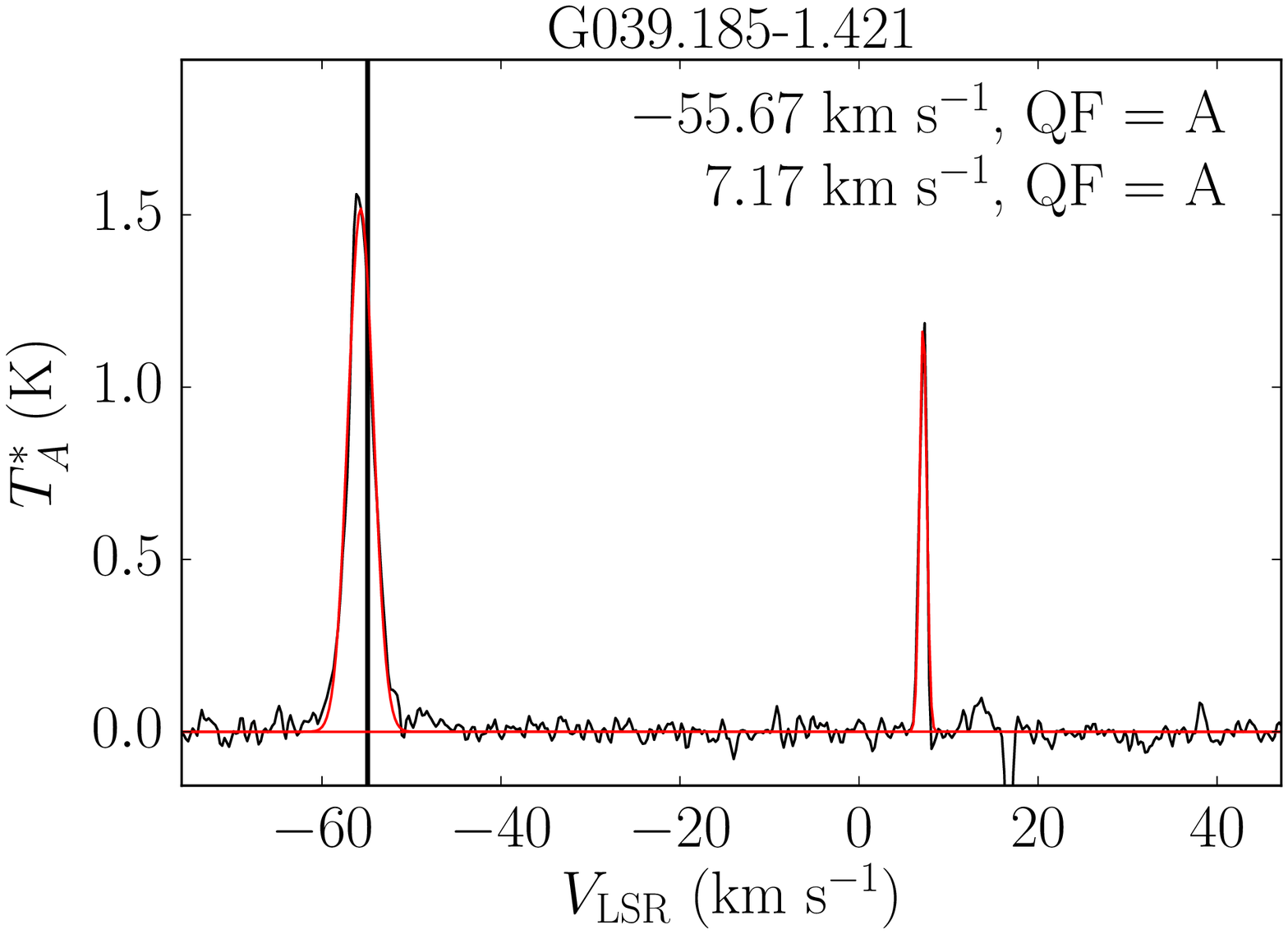}
\includegraphics[width=0.32\linewidth]{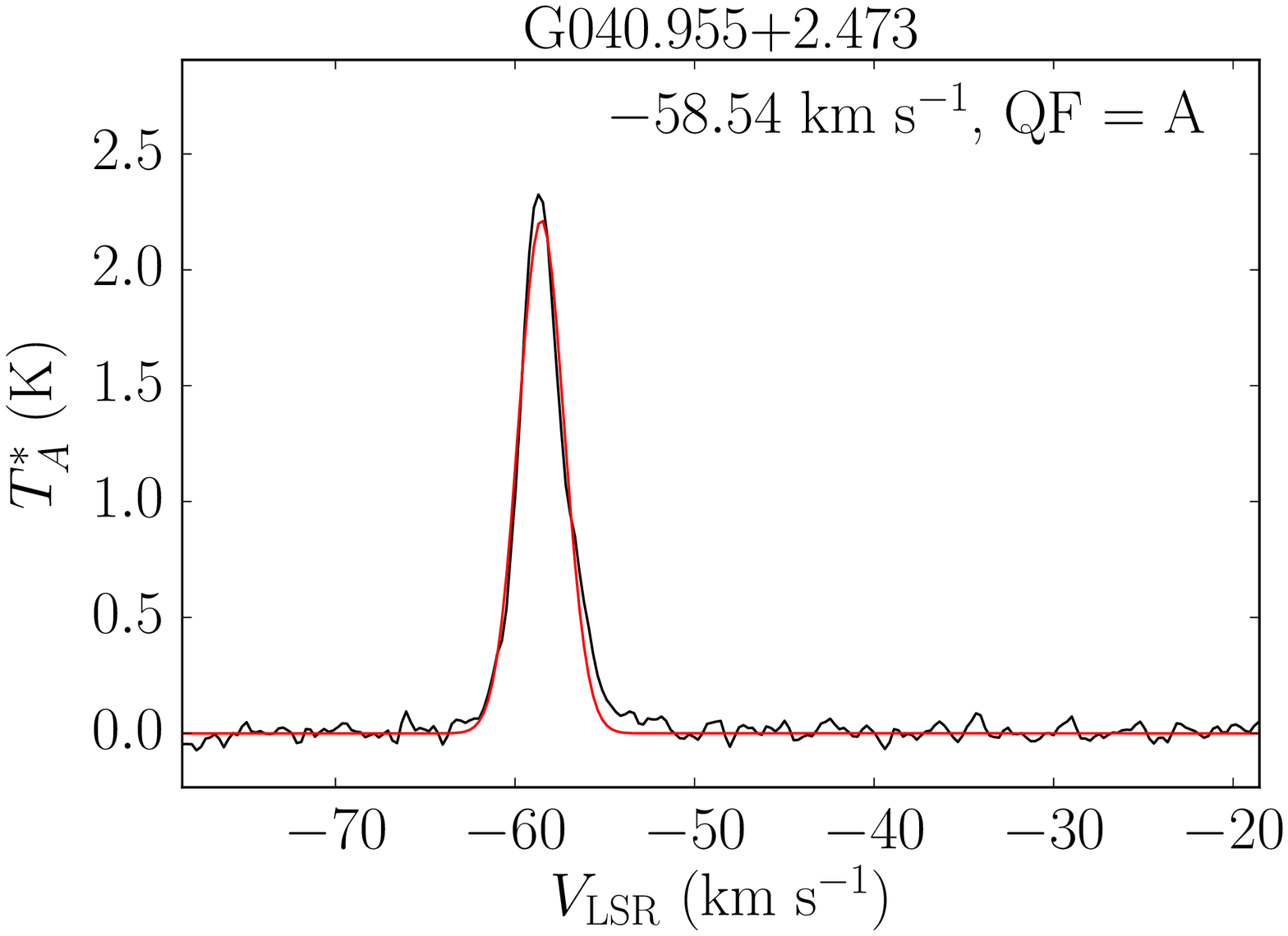}
\includegraphics[width=0.32\linewidth]{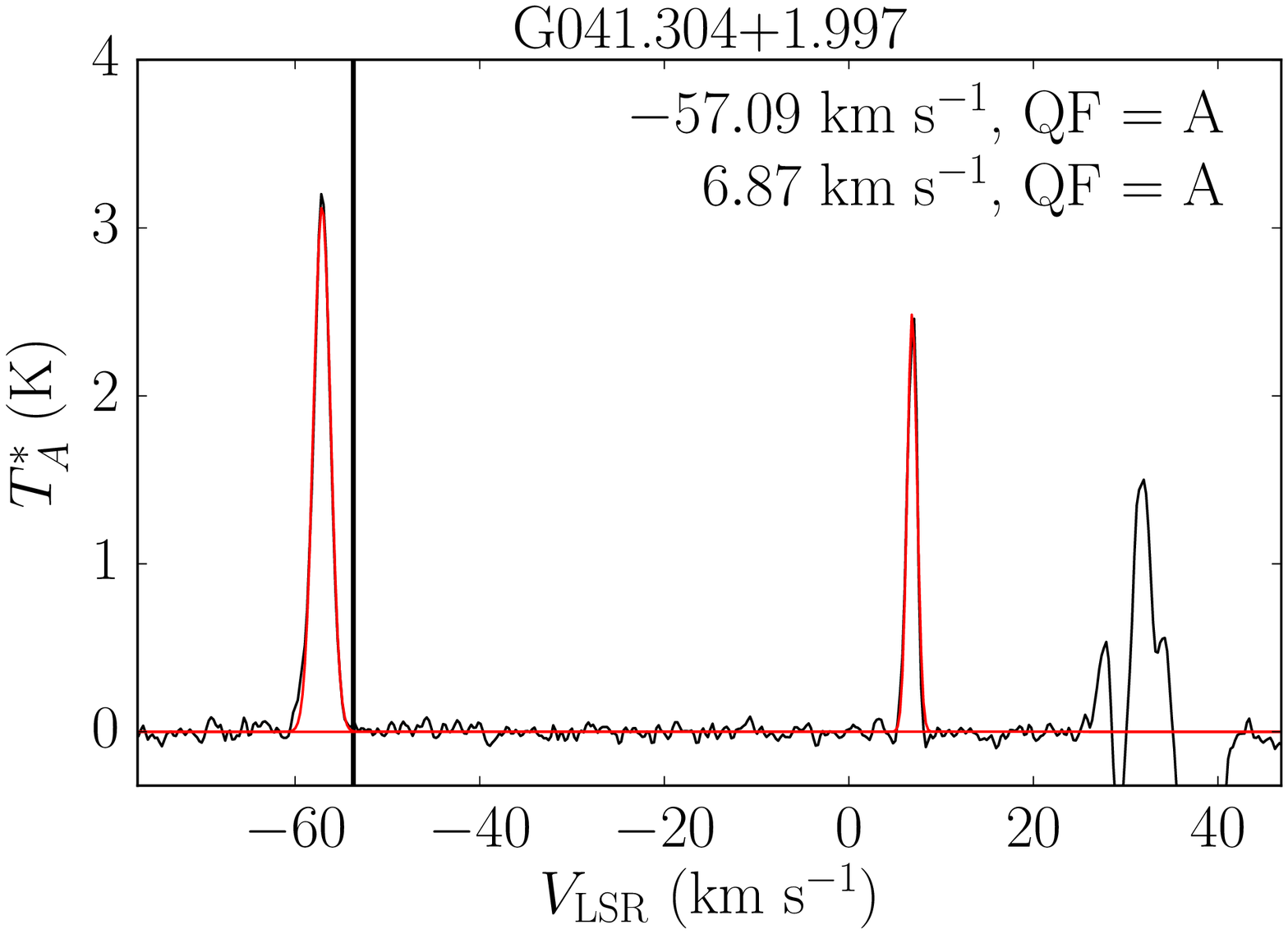}
\includegraphics[width=0.32\linewidth]{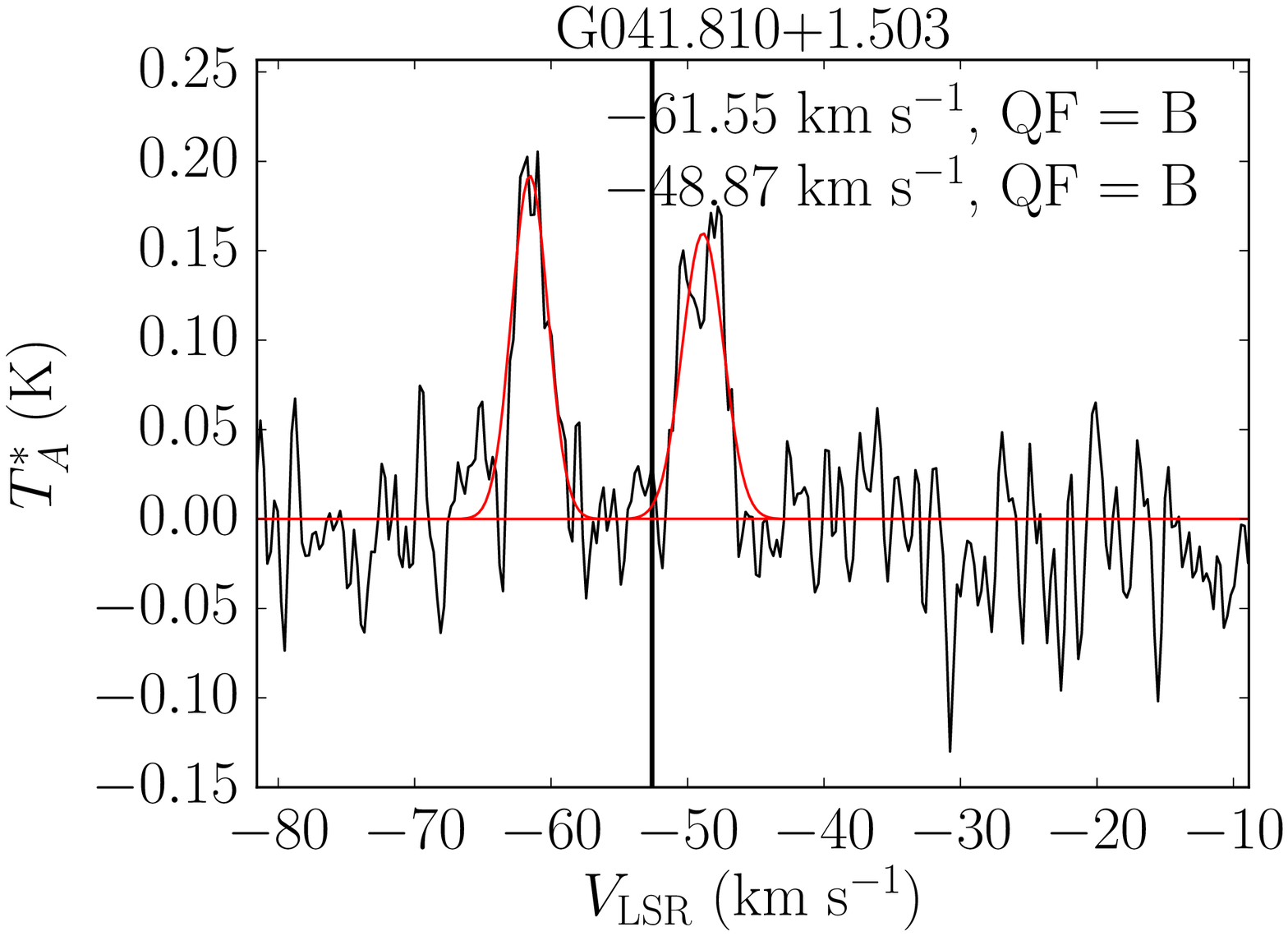}
\includegraphics[width=0.32\linewidth]{G042.210+1.081_12co.eps}
\includegraphics[width=0.32\linewidth]{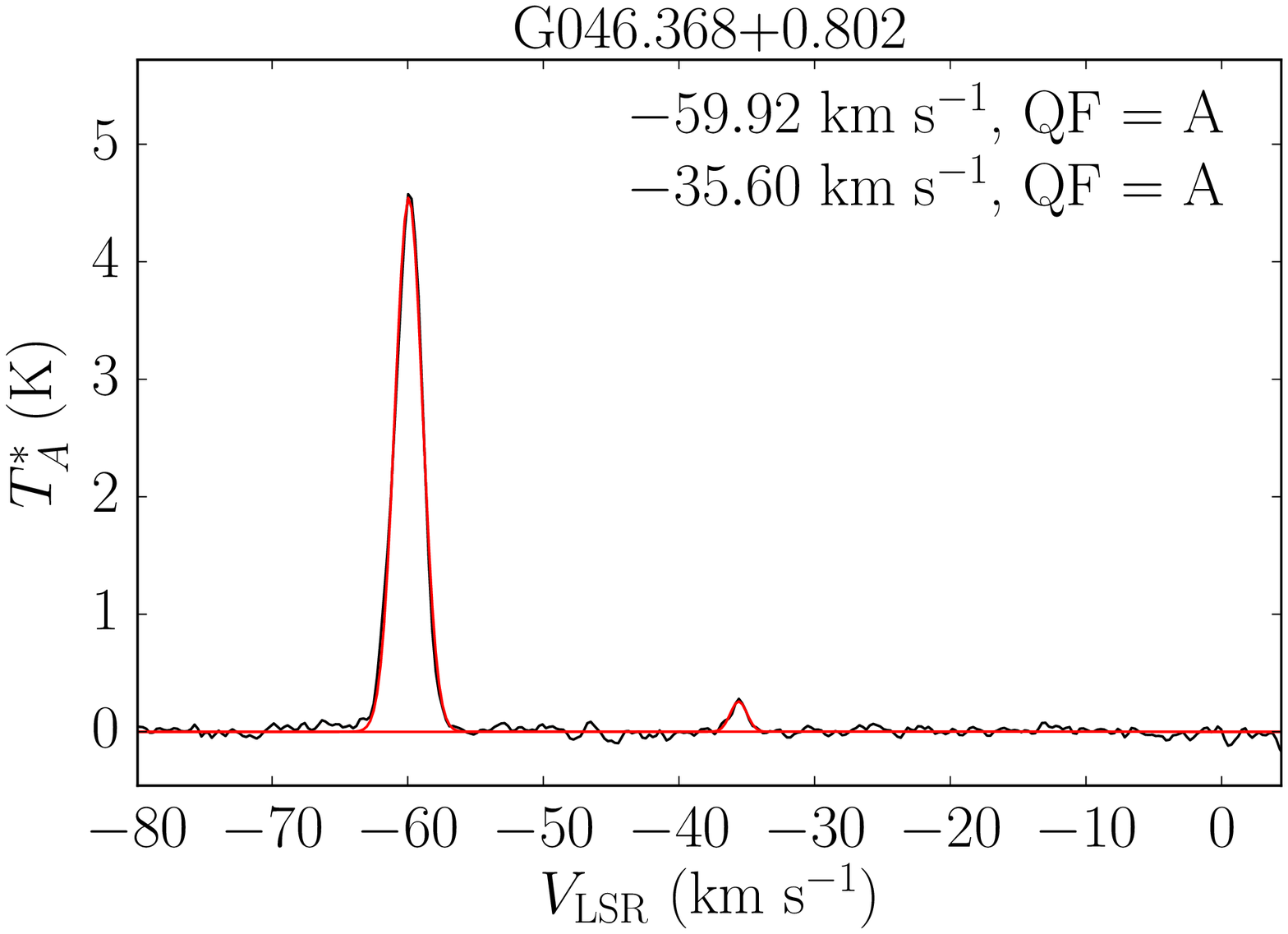}
\includegraphics[width=0.32\linewidth]{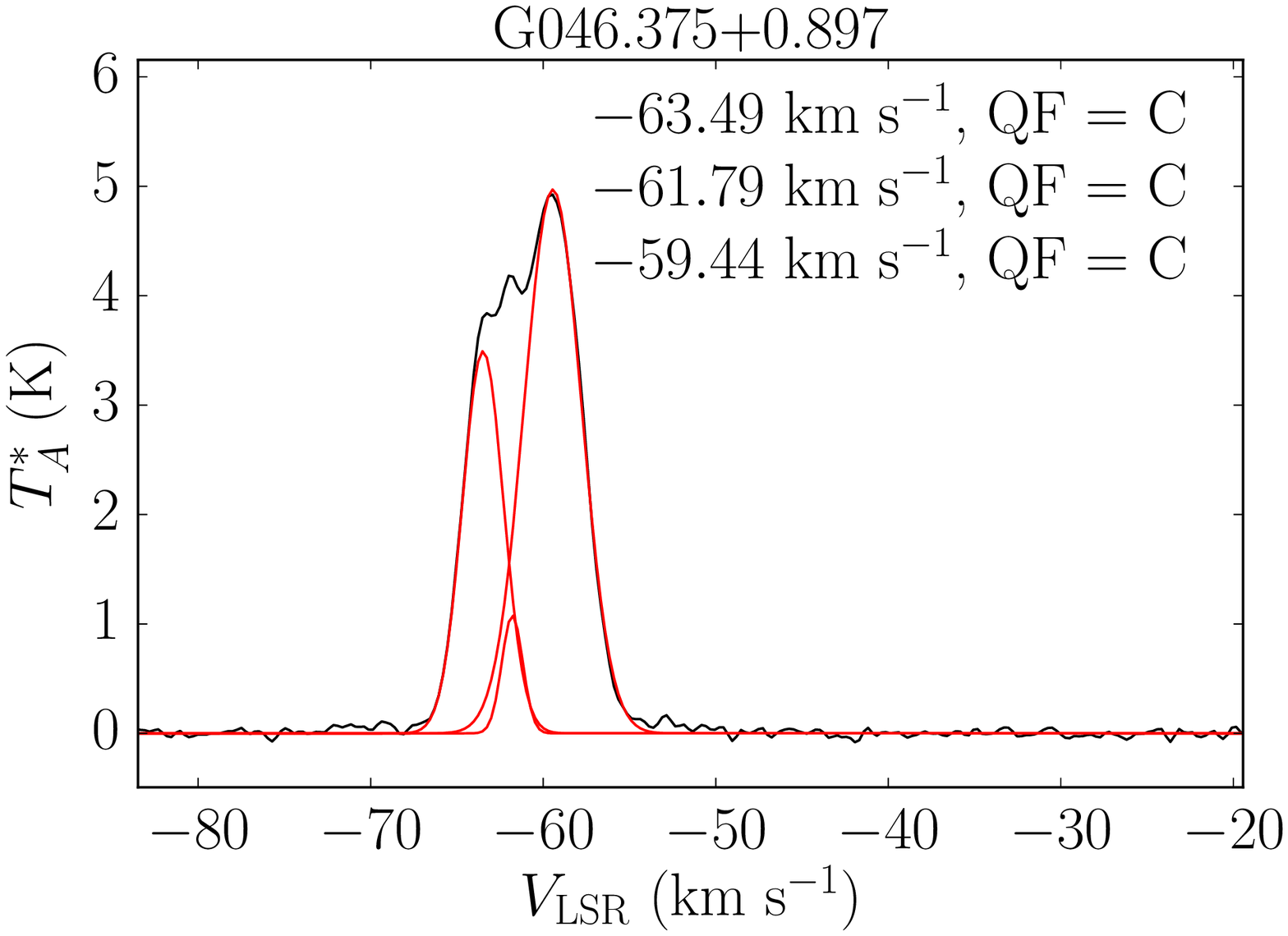}
\includegraphics[width=0.32\linewidth]{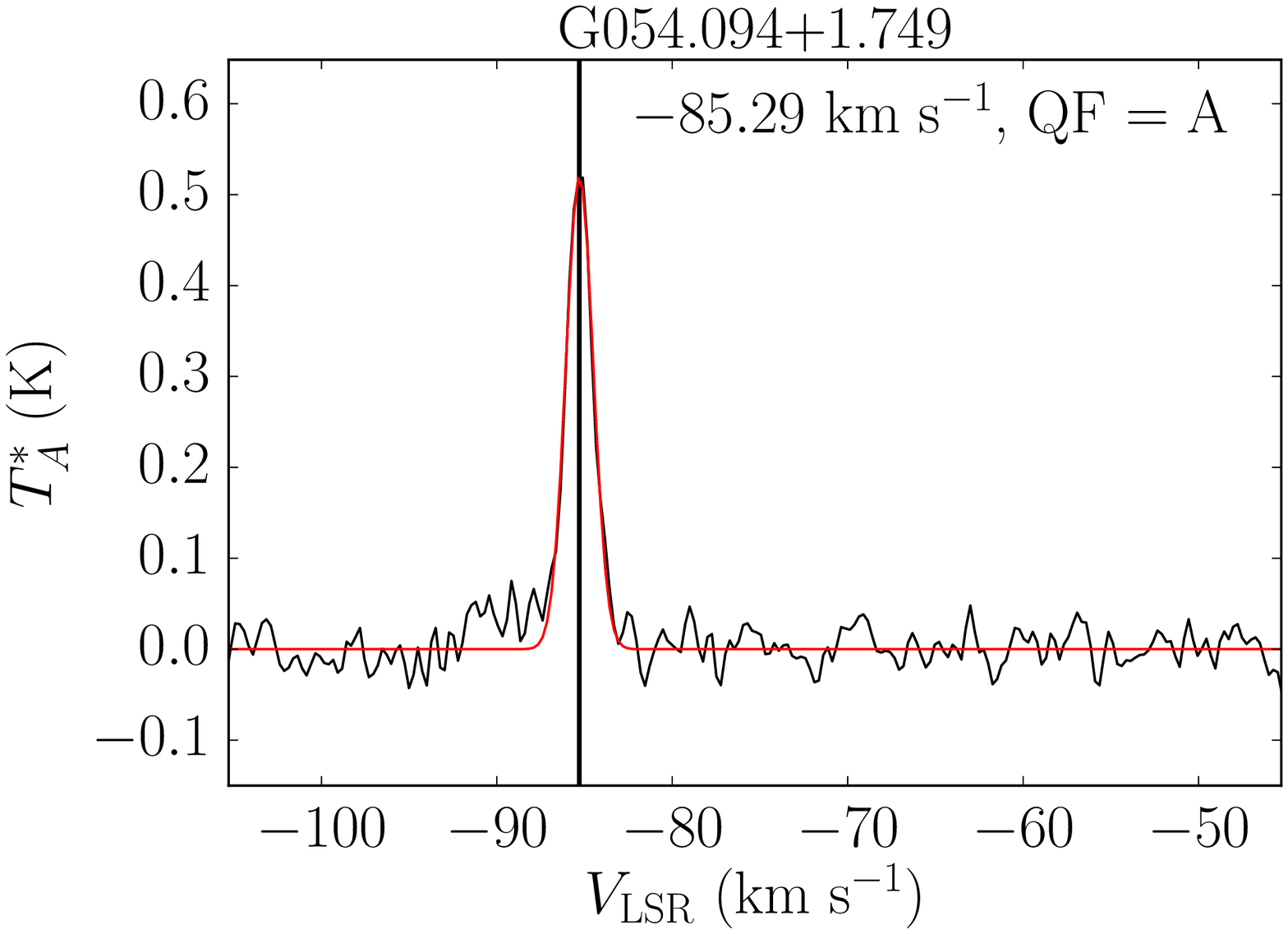}
\includegraphics[width=0.32\linewidth]{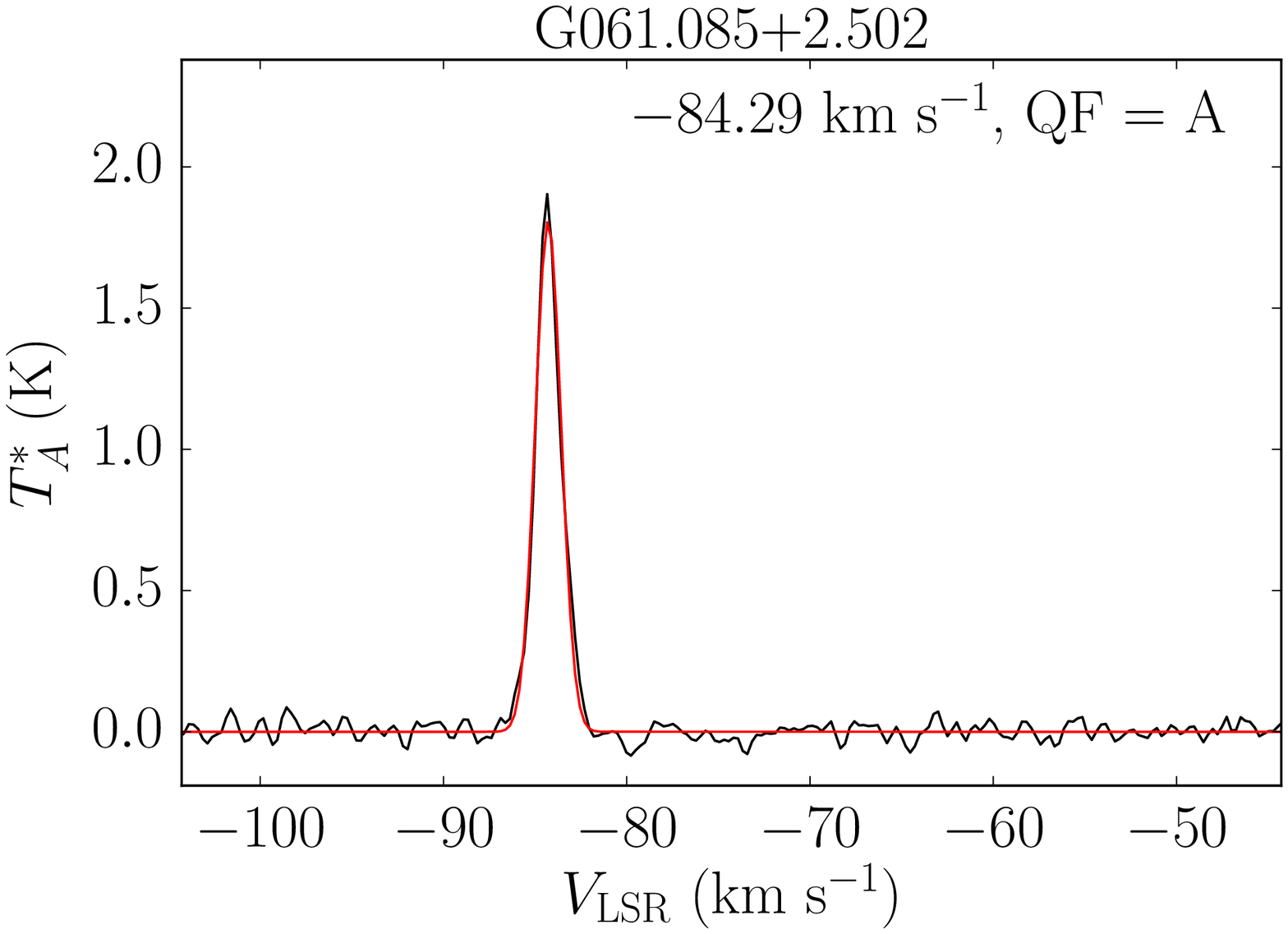}
\includegraphics[width=0.32\linewidth]{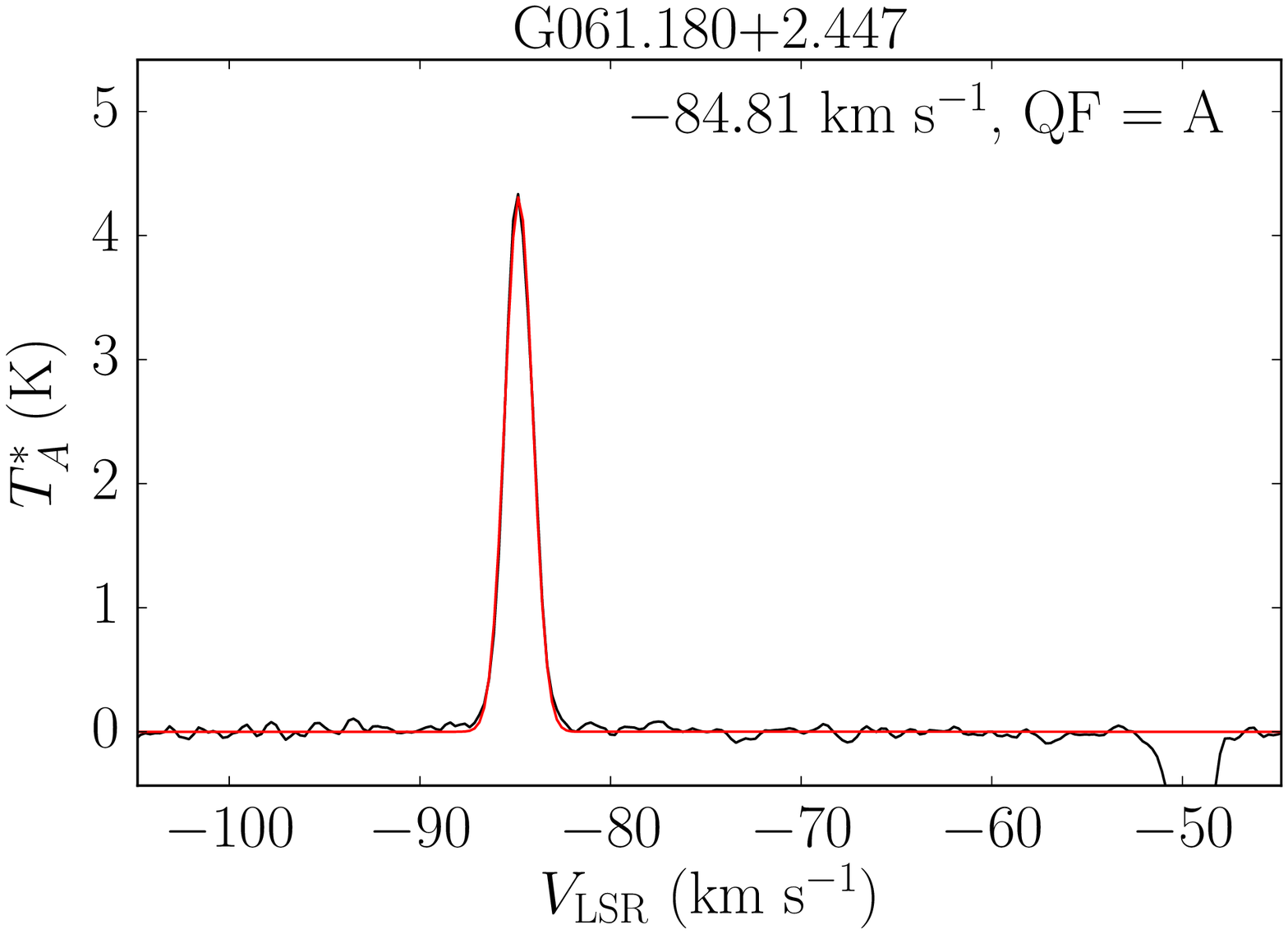}
\includegraphics[width=0.32\linewidth]{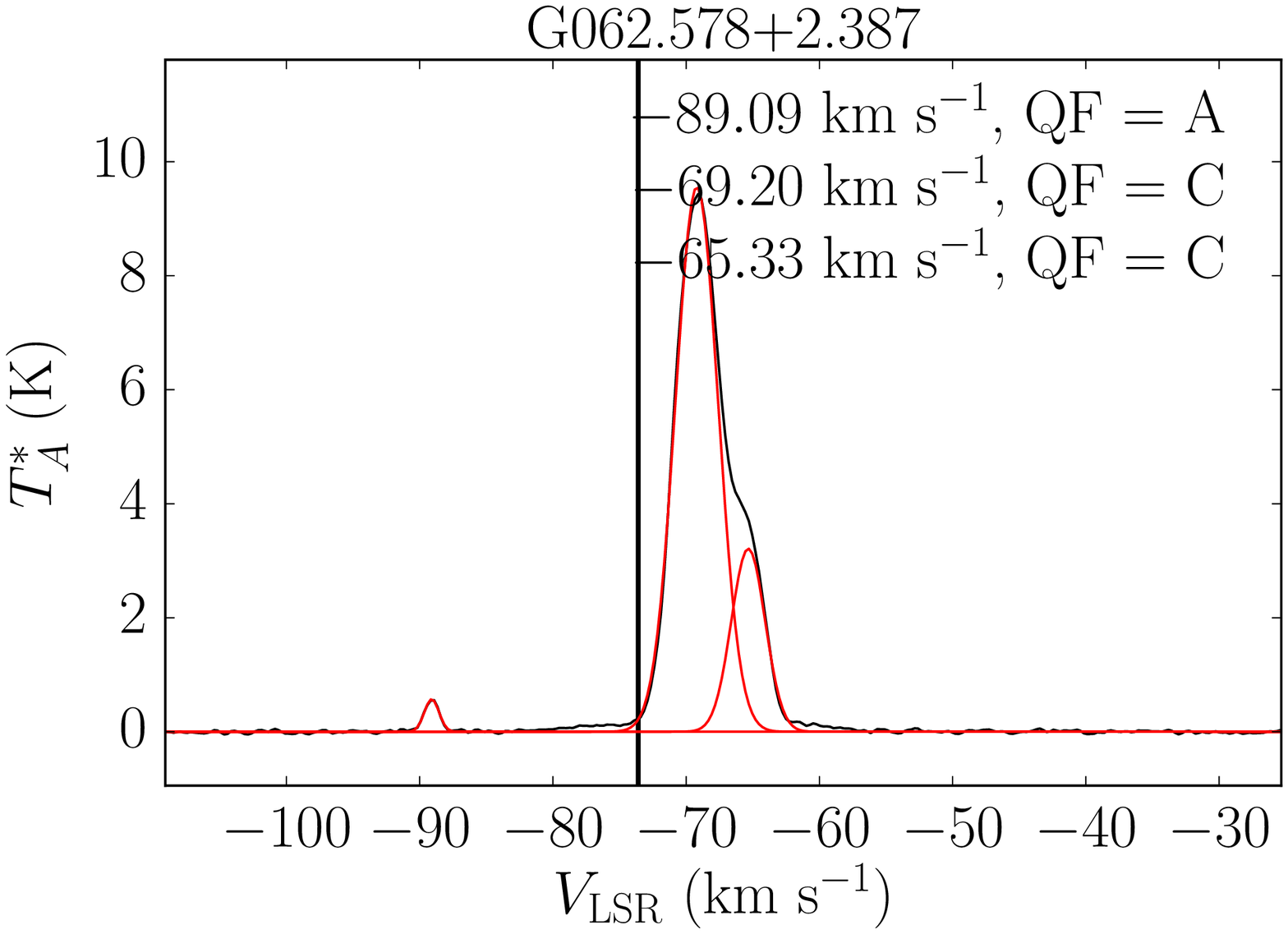}
\caption{\co\ spectra of emission lines originating within the OSC.
  Plotted is the antenna temperature as a function of the LSR
  velocity.  The black curves are the data and the red curves are
  Gaussian fits to the data.  The LSR velocity of each Gaussian
  profile, together with the QF, is shown in the right-hand corner of
  the plot. The vertical black line indicates the RRL velocity, when
  available.}
\label{fig:co12_osc}
\end{figure*}

\begin{figure*}[!htb]
\centering
\includegraphics[width=0.32\linewidth]{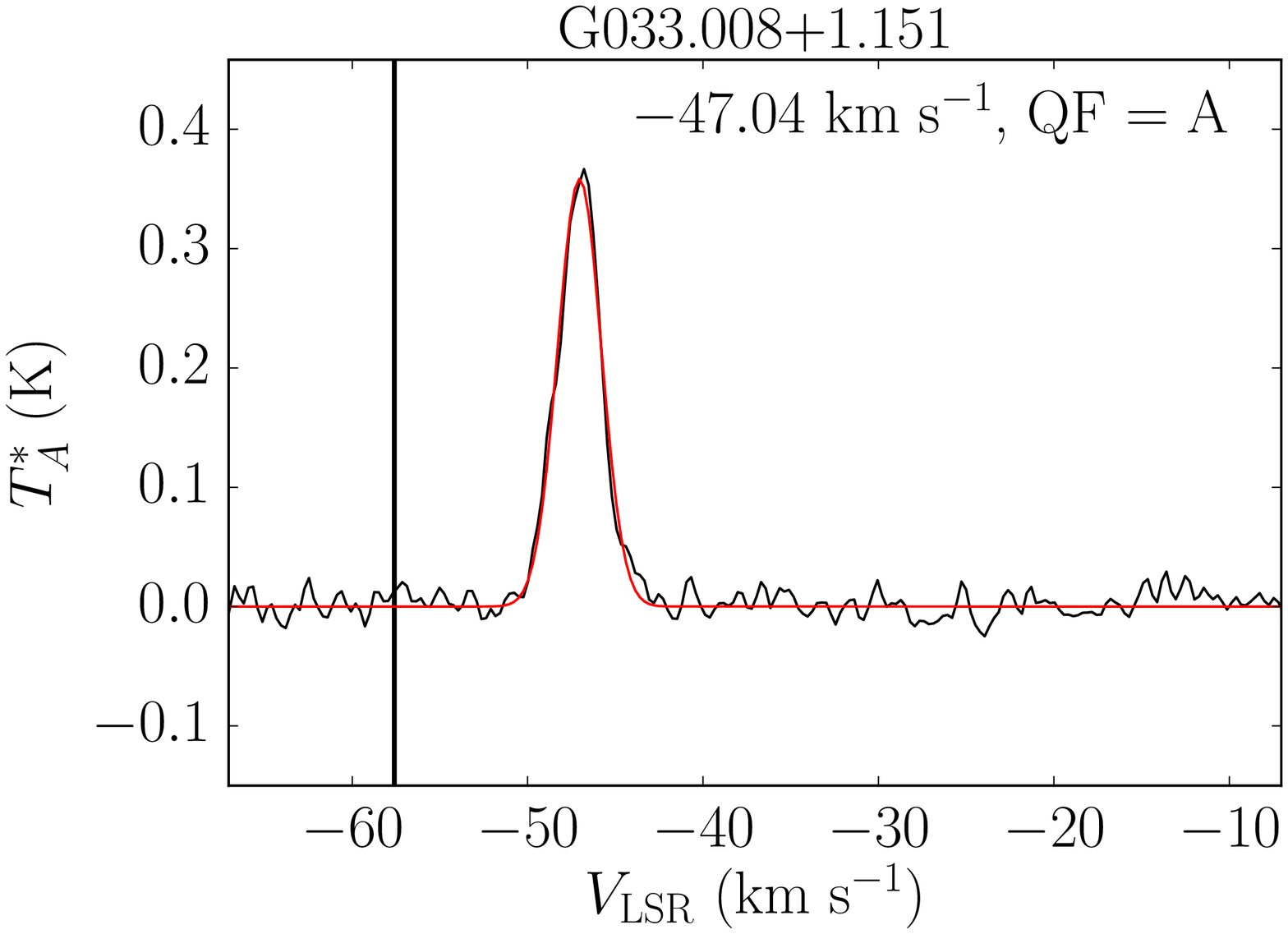}
\includegraphics[width=0.32\linewidth]{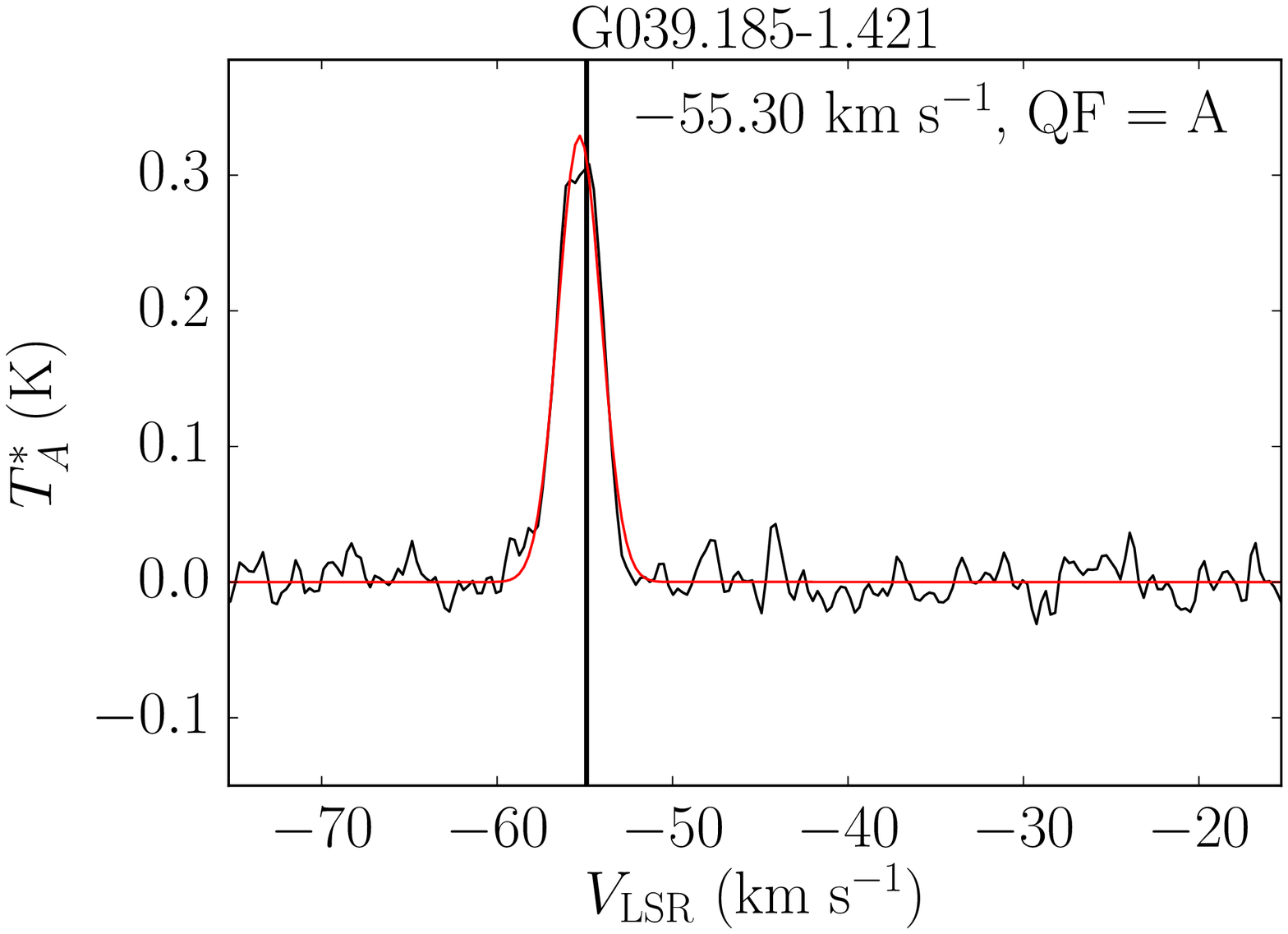}
\includegraphics[width=0.32\linewidth]{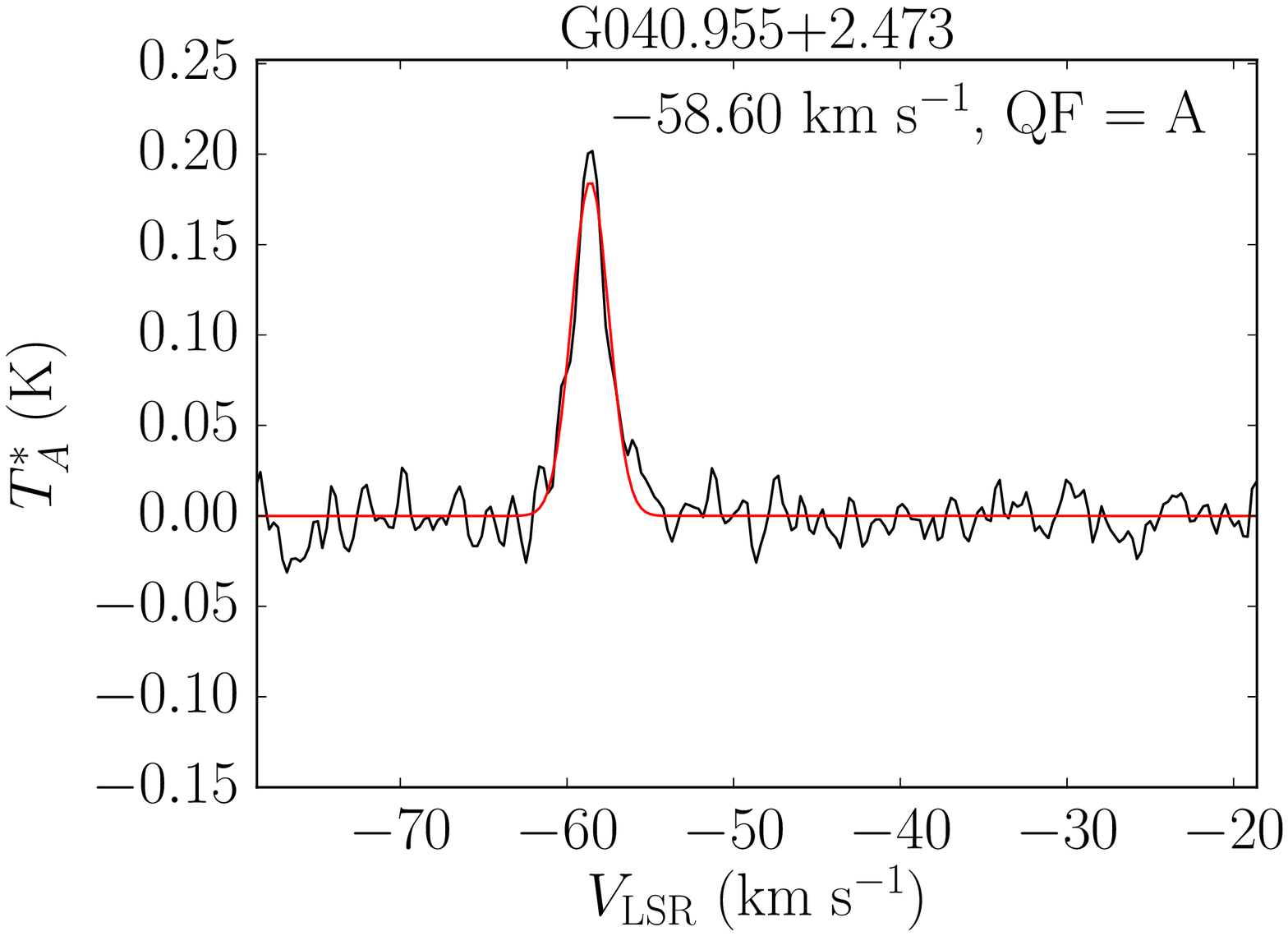}
\includegraphics[width=0.32\linewidth]{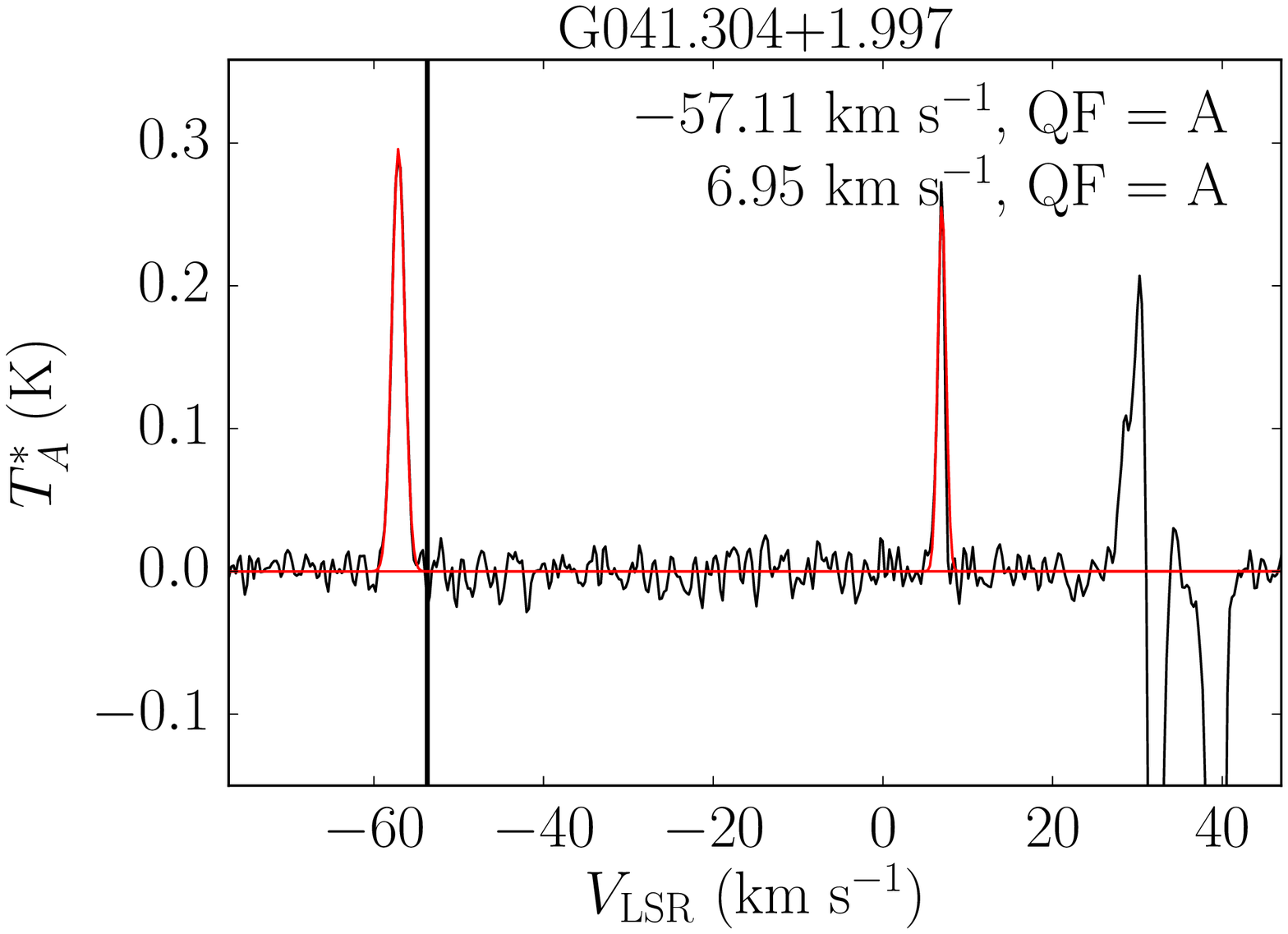}
\includegraphics[width=0.32\linewidth]{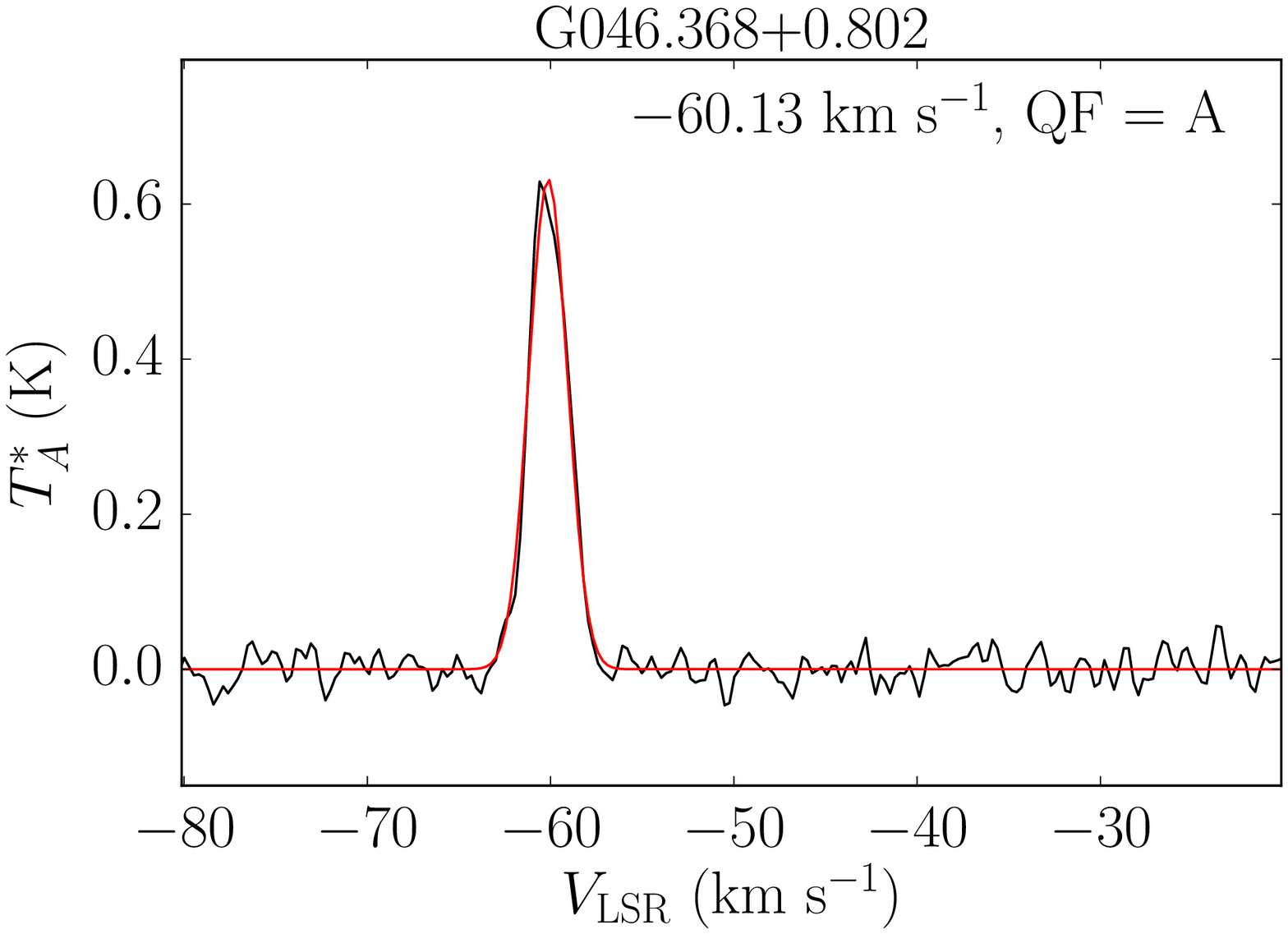}
\includegraphics[width=0.32\linewidth]{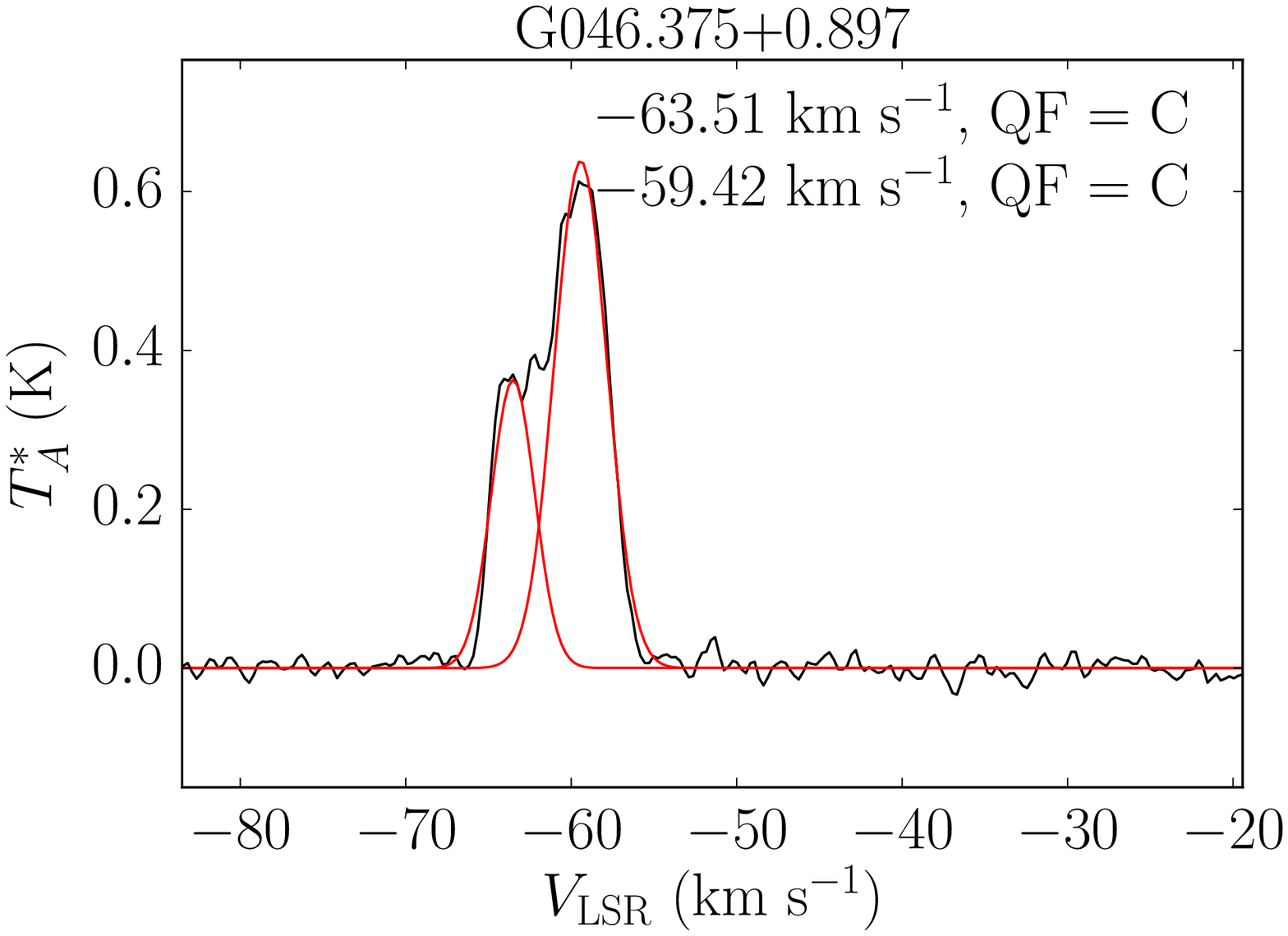}
\includegraphics[width=0.32\linewidth]{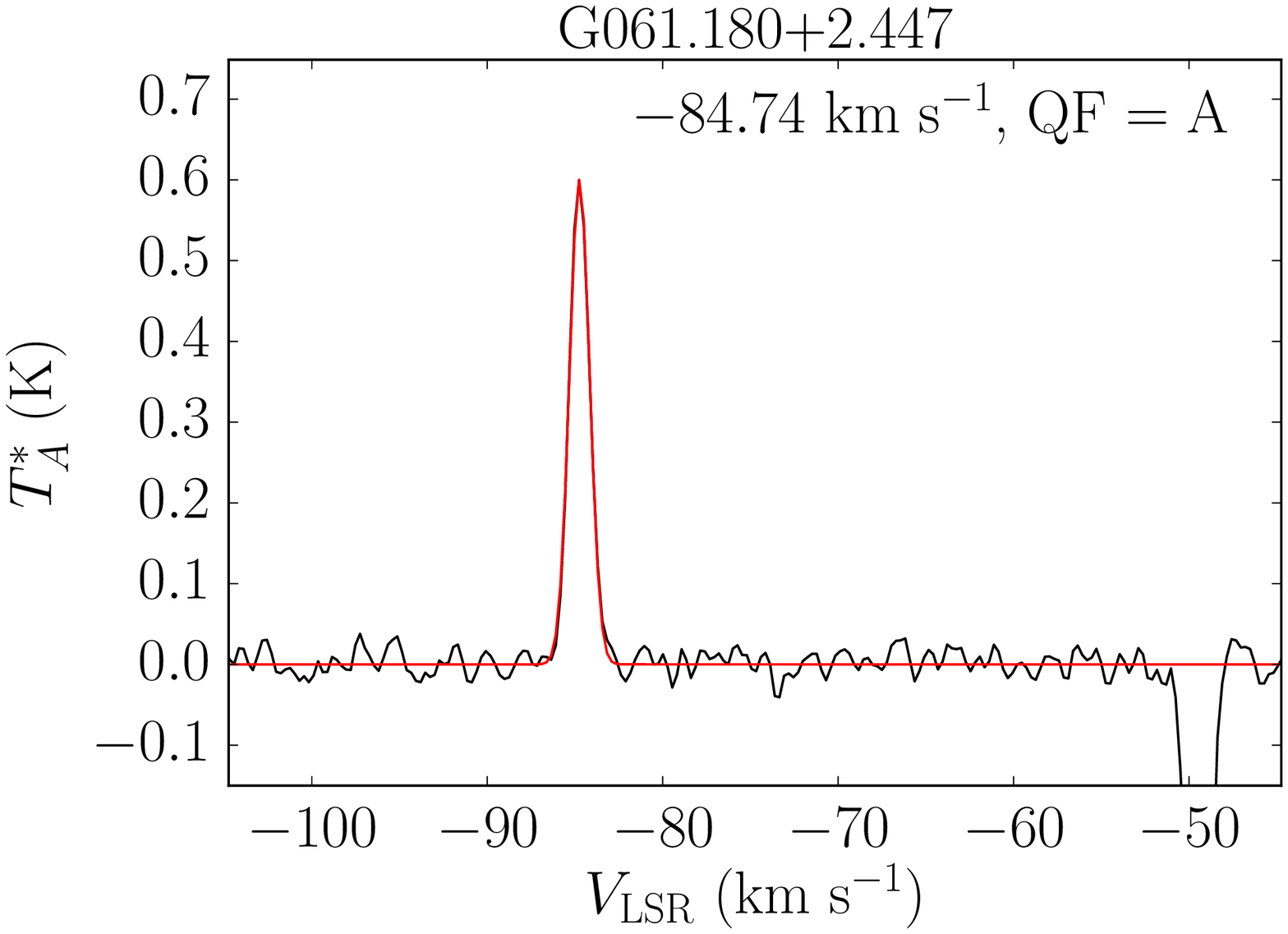}
\caption{\cor\ spectra of emission lines originating within the OSC.
  Plotted is the antenna temperature as a function of the LSR
  velocity.  The black curves are the data and the red curves are
  Gaussian fits to the data.  The LSR velocity of each Gaussian
  profile, together with the QF, is shown in the right-hand corner of
  the plot. The vertical black line indicates the RRL velocity, when
  available.}
\label{fig:co13_osc}
\end{figure*}

\section{Observations and Data Reduction}

We used the Arizona Radio Observatory (ARO) 12\m\ telescope to observe
the \co\ J=1--0 transition at 115.27120\ghz\ toward 78 \hii\ regions
and \hii\ region candidates located in the first Galactic quadrant. We
selected all \hii\ region candidates in the {\it WISE} Catalog of
Galactic \hii\ Regions \citep{anderson2012} that lay near the OSC
($\ell, b$) locus defined as \(b = 0.375^\circ + 0.075\times\ell\)
within the range \(20^\circ < \ell < 70^\circ\). Many of these targets
have measured radio continuum emission and are most likely bona fide
\hii\ regions \citep{armentrout2017}. In addition, we included 10
\hii\ regions from \citep{armentrout2017} with RRL velocities within
the OSC (\(\ell, V\)) locus defined as $V_{\rm LSR} = -1.6\, \ell \pm
15\kms$.  Since \co\ is often optically thick in Galactic molecular
clouds, we observed the optically thin \cor\ J=1--0 transition at
110.20132\ghz\ in a subset of targets with bright \co\ detections to
provide a more accurate measure of the molecular column density.

The ARO 12\m\ telescope is the European ALMA prototype antenna that
began operation on Kitt Peak in 2014.  The telescope's half-power
beam-width (HPBW) is 54\arcsec\ and 57\arcsec\ at 115\ghz\ and
110\ghz\ , respectively. The main beam efficiency is
\(\gsim90\%\) at these frequencies.  Our observations were performed
between February 11--16, 2016. The receiver consisted of the ALMA band
3, dual polarization, sideband-separating mixers with typical on-sky
system temperatures of $\sim 300$\K. We employed both the filter bank
spectrometer with 256 channels at 2\mhz\ spectral resolution
(5\kms\ at 115\ghz), and the millimeter autocorrelator (MAC) with 4096
channels at 195\khz\ spectral resolution (0.5\kms\ at 115\ghz).  Both
spectrometers accept two intermediate frequency signals consisting of
the two orthogonal polarizations.  Here we only consider data from the
MAC spectrometer given the narrow line widths of CO ($\sim 1$\kms).
The typical atmospheric optical depth at zenith was \(\tau_0 \sim
0.2\).

We made total power, position switched observations where the
reference position (Off) is offset by 20\arcmin\ in azimuth from the
source (On).  The On and Off positions were observed for 5 minutes
each with a switching rate of 30 seconds, for a total time of 10
minutes.  Typically 1--2 total power pairs were sufficient to detect
\co emission but in some cases we integrated longer.  The telescope
pointing and focus were corrected every 1--2 hours by peaking on
Jupiter, Venus, or Saturn.  The typical pointing accuracy was $\sim
2$\arcsec\ rms.  At the start of each session we checked the tuning of
the spectrometers by observing the test source M17SW.

The data were reduced and analyzed using TMBIDL, an IDL single-dish
software package \citep{bania2016}.\footnote{V7.1, see
  \url{https://github.com/tvwenger/tmbidl}} The data reduction and
analysis were performed independently by three of the authors.  Each
spectrum was visually inspected. We discarded \({\sim}1\%\) of the
spectra due to poor baseline structure.  For each target the data were
averaged over all total power pairs and polarizations to produce a
single, averaged spectrum.  We modeled the spectral baselines with a
third-order polynomial function that was subtracted from the data to
remove any sky continuum emission or residual baseline structure in
the spectrum.  We fitted a Gaussian function to each profile using a
Levenberg-Markwardt least squares method \citep{markwardt2009} to
derive the peak intensity, the full width at half-maximum (FWHM) line
width, and the local standard of rest (LSR) velocity.

We compared the results of each independent analysis. The results were
similar for \({\sim}90\%\) of the spectra.  Any differences typically
involved how best to fit complex, blended profiles.  Multiple
\co\ emission components were detected in about 2/3 of the targets.
These directions thus contain several \co\ clouds along the
line-of-sight.  Furthermore, because CO is pervasive in the inner
Galactic plane, there was often \co\ emission detected in the Off
position. This creates apparent absorption lines in the processed
position-switched spectrum. Since any emission components near these
absorption features are contaminated by this Off emission, we
eliminated these components from any further analysis. The
\cor\ spectra were analyzed in the same manner.

\begin{figure*}[!htb]
\centering
\includegraphics[width=\linewidth]{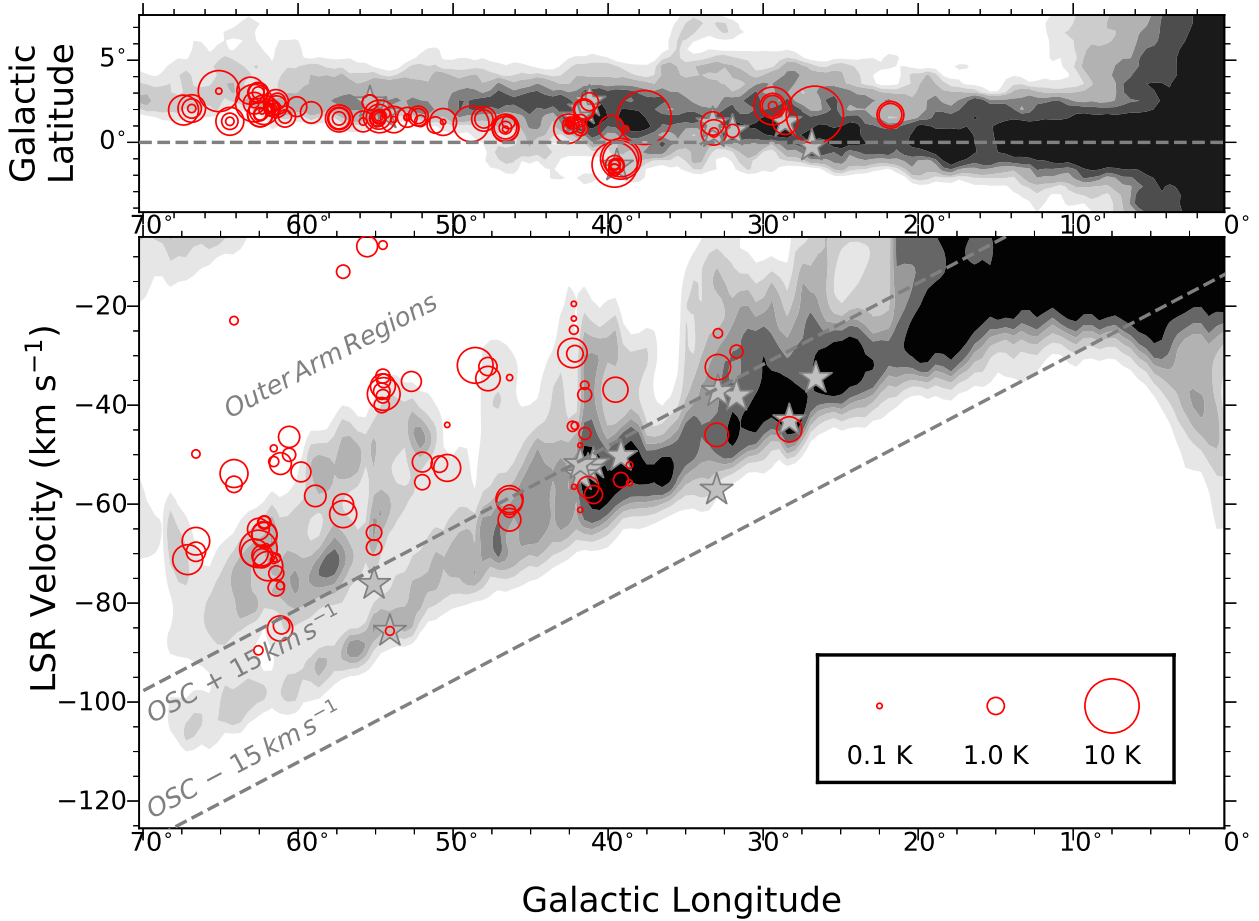}
\caption{The outer Scutum-Centaurus spiral arm as traced by integrated
  \hi\ emission, CO emission, and \hii\ regions. Top:
  velocity-integrated \hi\ emission (grayscale image) tracing the OSC
  spiral arm, summed over a 14\kms\ wide window following the center
  velocity given by $V_{\rm LSR} = -1.6\kms\,{\rm deg}^{-1} \times
  \ell$.  CO emission detected here toward {\it WISE} \hii\ regions is
  shown as red circles with sizes scaled by the \co\ peak line
  intensity.  Gray stars correspond to \hii\ regions consistent with
  the location of the OSC \citep{armentrout2017}.  Bottom:
  longitude-velocity diagram of \hi\ emission (grayscale image),
  summed over a 3\degper 5 window following the arm in latitude with
  $b = $0\degper 75 + 0.75 $\times \ell$.  The symbols are the same as
  in the top panel.  The OSC locus is defined by $V_{\rm LSR} = -1.6\,
  \ell \pm 15$\kms\ and shown by the dashed lines. (Reproduced from
  \citet{dame2011} and \citet{armentrout2017}.)}
\label{fig:lv_diagram}
\end{figure*}

We assigned a Quality Factor (A, B, or C) to every CO emission
line. This judgment was based on the signal-to-noise ratio and
expected line properties (e.g., the line widths should be a few
\(\kms\) at most). A quality factor of ``A'' was given to profiles
with a signal-to-noise ratio larger than 10 with no blending and flat
spectral baselines.  Targets with either lower signal-to-noise ratio
or some blending of lines were given a quality factor ``B''.  Here the
blending should not be so severe that the individual peaks could not
be visually detected.  A quality factor of ``C'' was given to the
remaining targets which were often blended and more difficult to fit.
A quality factor of D was assigned to spectra with no CO detections.

\section{Results and Discussion}

We detect CO toward $\sim 80$\% of our targets. About 2/3 of our
targets have at least one CO emission line at negative velocity,
placing it beyond the Solar orbit.  The results are summarized in
Figures~\ref{fig:qf}--\ref{fig:co13_osc} and
Tables~\ref{tab:co12}--\ref{tab:co13}.  We measure 117 distinct
\co\ components toward 62 of 78 targets, and 40 \cor\ components
toward 27 of 30 targets. Representative \co\ spectra that span the
range of quality factors A--C are shown in Figure~\ref{fig:qf}.  All
line intensities are in units of the antenna temperature corrected for
atmospheric attenuation, radiative loss, and rearward scattering and
spillover, $T_{\rm A}^{*}$.  The Gaussian fit parameters are shown in
Table~\ref{tab:co12} and Table~\ref{tab:co13} for the \co\ and
\cor\ spectra, respectively.  Listed are the target name, the Gaussian
fit parameters, the root-mean-square noise in the line-free baseline,
RMS, the total integration time, $t_{\rm intg}$, and the quality
factor, QF.  The Gaussian fits include the peak line intensity,
$T_{\rm L}$, the LSR velocity, $V_{\rm LSR}$, and the FWHM line width,
$\Delta{V}$, and their associated 1-$\sigma$ uncertainties.  For
\co\ spectra with multiple emission components, in
Table~\ref{tab:co12} we append a letter at the end of the target name
(a, b, c, etc.) in order of decreasing line intensity.

In most cases we can associate each \cor\ component with a
corresponding \co\ component using the LSR velocity (i.e. the \co\ and
\cor\ components have the same LSR velocity, within \(2\kms\)). There
are four exceptions.  In G050.901+1.056 and G062.194+1.715 there are
two \cor\ components that correspond to one \co\ component (identified
as G050.901+1.056a, G050.901+1.056b, G062.194+1.715a1, and
G062.194+1.715a2 in Table~\ref{tab:co13}). The reverse occurs in
G066.608+2.061 where there is one \cor\ component corresponding to two
\co\ components (identified as G066.608+2.061ab in
Table~\ref{tab:co13}). These differences may be caused by opacity
effects where the \cor\ emission is arising from a region that is
optically thick in \co. This can cause self-absorption in the
\co\ spectrum. The signal-to-noise ratio will also affect the ability
to resolve multiple components. Finally, for G061.154+2.170 there are
\co\ components at velocities where the \cor\ emission is corrupted by
emission in the Off position. The \co\ and \cor\ spectra for these
four targets are shown in Figure~\ref{fig:special}.

Following \citet{armentrout2017}, we deem molecular emission
components to be located in the OSC whenever the molecular line LSR
velocity is within the ($\ell, v$) locus defined by $V_{\rm LSR} =
-1.6\, \ell \pm 15$\kms.  We find that 17 \co\ emission lines toward
14 targets and 8 \cor\ emission lines toward 7 targets originate
within the OSC. Of these detections, 7 \co\ targets (G028.319+1.243,
G033.008+1.151, G039.185-1.421, G041.304+1.997, G041.810+1.503,
G054.094+1.749, G062.578+2.387) and 3 \cor\ targets (G033.008+1.151,
G039.185-1.421, G041.304+1.997) have a measured RRL velocity in the
\textit{WISE} Catalog of Galactic \hii\ regions \citep{anderson2012}.
We show spectra for all of our OSC detections in
Figure~\ref{fig:co12_osc} and \ref{fig:co13_osc} for \co\ and \cor,
respectively. Two OSC targets have multiple \co\ or \cor\ spectral
line components that lie within the OSC: G038.627+0.813 and
G046.375+0.897.  Our OSC targets may not all be unique objects.  There
are two target pairs --- G046.368+0.802/G046.375+0.897 and
G061.085+2.502/G061.180+2.447 --- that lie near the same ($\ell, b,
v$) locus and therefore may be part of the same star formation
complex.  This may account for the multiple component spectrum seen
toward G046.375+0.897.  Finally, six of our OSC targets are probably
related to sources detected by \citet{sun2017} in CO emission:
G038.627+0.813ab (MWISP G38.533+0.892), G039.185$-$1.421 (MWISP
G39.175$-$1.425), G040.955+2.473 (MWISP G40.958+2.483), G041.304+1.997
(MWISP G41.308+2.000), G041.810+1.503 (MWISP G41.758+1.567, MWISP
G41.733+1.517, MWISP G41.742+1.458), and G042.210+1.081 (MWISP
G42.192+1.083).

\startlongtable
\begin{deluxetable*}{lrrrrrrrrr}
\tabletypesize{\scriptsize}
\tablecaption{\co Emission Line Parameters\label{tab:co12}}
\tablehead{
\colhead{Name} & 
\colhead{$T_L$} & 
\colhead{$\sigma_{T_L}$} & 
\colhead{$V_{\rm LSR}$} & 
\colhead{$\sigma_{V_{\rm LSR}}$} & 
\colhead{$\Delta V$} & 
\colhead{$\sigma_{\Delta V}$} & 
\colhead{RMS} & 
\colhead{$t_{\rm intg}$} & 
\colhead{QF}\\
\colhead{} & 
\colhead{(K)} & 
\colhead{(K)} & 
\colhead{(\kms)} & 
\colhead{(\kms)} & 
\colhead{(\kms)} & 
\colhead{(\kms)} & 
\colhead{(K)} & 
\colhead{(min.)} & 
\colhead{}
}
\startdata
G021.541+1.676a
 & $
      4.91$ & $
      0.09$ & $
     31.07$ & $
      0.04$ & $
      2.33$ & $
      0.09$ & $
    0.050$ & $
 19.5$ & 
B
 \\ 
G021.541+1.676b
 & $
      3.56$ & $
      0.09$ & $
     33.66$ & $
      0.06$ & $
      2.28$ & $
      0.12$ & $
    0.050$ & $
 19.5$ & 
B
 \\ 
G023.610$-$0.212
 & \nodata & \nodata & \nodata & \nodata & \nodata & \nodata & 
0.034 & 
 39.0 & 
D
 \\ 
G026.380+1.678
 & \nodata & \nodata & \nodata & \nodata & \nodata & \nodata & 
0.052 & 
 19.5 & 
D
 \\ 
G026.418+1.683
 & $
     22.30$ & $
      0.17$ & $
     43.32$ & $
      0.02$ & $
      4.49$ & $
      0.04$ & $
    0.031$ & $
 38.9$ & 
A
 \\ 
G026.964+1.644
 & \nodata & \nodata & \nodata & \nodata & \nodata & \nodata & 
0.046 & 
 19.5 & 
D
 \\ 
\textbf{G028.319+1.243}
 & $
      4.32$ & $
      0.04$ & $
    -45.75$ & $
      0.02$ & $
      3.30$ & $
      0.04$ & $
    0.046$ & $
 19.6$ & 
A
 \\ 
G029.136+2.219a
 & $
      9.73$ & $
      0.13$ & $
     32.88$ & $
      0.03$ & $
      2.80$ & $
      0.07$ & $
    0.042$ & $
 19.4$ & 
B
 \\ 
G029.136+2.219b
 & $
      4.30$ & $
      0.13$ & $
     36.37$ & $
      0.07$ & $
      2.85$ & $
      0.20$ & $
    0.042$ & $
 19.4$ & 
B
 \\ 
G029.136+2.219c
 & $
      2.99$ & $
      0.16$ & $
      9.57$ & $
      0.04$ & $
      1.64$ & $
      0.10$ & $
    0.042$ & $
 19.4$ & 
C
 \\ 
G029.136+2.219d
 & $
      0.50$ & $
      0.19$ & $
      6.19$ & $
      0.21$ & $
      1.17$ & $
      0.51$ & $
    0.042$ & $
 19.4$ & 
C
 \\ 
G031.727+0.698
 & $
      1.16$ & $
      0.03$ & $
    -30.50$ & $
      0.03$ & $
      2.60$ & $
      0.07$ & $
    0.030$ & $
 39.0$ & 
A
 \\ 
G032.928+0.606a
 & $
      4.39$ & $
      0.02$ & $
    -33.55$ & $
      0.01$ & $
      3.89$ & $
      0.03$ & $
    0.043$ & $
 19.5$ & 
B
 \\ 
G032.928+0.606b
 & $
      0.58$ & $
      0.02$ & $
    -26.91$ & $
      0.07$ & $
      3.41$ & $
      0.17$ & $
    0.043$ & $
 19.5$ & 
B
 \\ 
\textbf{G033.008+1.151}
 & $
      3.75$ & $
      0.05$ & $
    -46.84$ & $
      0.02$ & $
      2.79$ & $
      0.05$ & $
    0.042$ & $
 19.5$ & 
A
 \\ 
G034.172+1.055
 & \nodata & \nodata & \nodata & \nodata & \nodata & \nodata & 
0.039 & 
 19.5 & 
D
 \\ 
G037.419+1.514
 & $
     19.96$ & $
      0.34$ & $
     44.19$ & $
      0.03$ & $
      3.47$ & $
      0.07$ & $
    0.028$ & $
 38.9$ & 
C
 \\ 
\textbf{G038.627+0.813a}
 & $
      0.30$ & $
      0.01$ & $
    -52.76$ & $
      0.05$ & $
      2.55$ & $
      0.13$ & $
    0.028$ & $
 38.9$ & 
B
 \\ 
\textbf{G038.627+0.813b}
 & $
      0.24$ & $
      0.01$ & $
    -56.25$ & $
      0.06$ & $
      2.28$ & $
      0.15$ & $
    0.028$ & $
 38.9$ & 
B
 \\ 
G038.931$-$1.003a
 & $
     11.27$ & $
      0.30$ & $
     51.60$ & $
      0.05$ & $
      2.10$ & $
      0.14$ & $
    0.035$ & $
 19.5$ & 
C
 \\ 
G038.931$-$1.003b
 & $
      8.83$ & $
      0.71$ & $
     49.93$ & $
      0.06$ & $
      1.74$ & $
      0.07$ & $
    0.035$ & $
 19.5$ & 
C
 \\ 
\textbf{G039.185$-$1.421a}
 & $
      1.52$ & $
      0.01$ & $
    -55.67$ & $
      0.01$ & $
      3.32$ & $
      0.04$ & $
    0.027$ & $
 34.1$ & 
A
 \\ 
G039.185$-$1.421b
 & $
      1.19$ & $
      0.03$ & $
      7.17$ & $
      0.01$ & $
      0.93$ & $
      0.02$ & $
    0.027$ & $
 34.1$ & 
A
 \\ 
G039.328$-$1.350a
 & $
     13.97$ & $
      0.07$ & $
     53.35$ & $
      0.01$ & $
      2.19$ & $
      0.01$ & $
    0.034$ & $
 19.5$ & 
B
 \\ 
G039.328$-$1.350b
 & $
      2.25$ & $
      0.08$ & $
     72.49$ & $
      0.03$ & $
      1.99$ & $
      0.08$ & $
    0.034$ & $
 19.5$ & 
A
 \\ 
G039.328$-$1.350c
 & $
      0.90$ & $
      0.07$ & $
     48.53$ & $
      0.08$ & $
      2.19$ & $
      0.20$ & $
    0.034$ & $
 19.5$ & 
A
 \\ 
G039.328$-$1.350d
 & $
      0.60$ & $
      0.13$ & $
      7.61$ & $
      0.07$ & $
      0.66$ & $
      0.16$ & $
    0.034$ & $
 19.5$ & 
B
 \\ 
G039.342$-$1.676
 & $
      0.92$ & $
      0.02$ & $
     56.17$ & $
      0.02$ & $
      1.56$ & $
      0.04$ & $
    0.034$ & $
 19.5$ & 
A
 \\ 
G039.343$-$1.675
 & $
      0.94$ & $
      0.01$ & $
     56.17$ & $
      0.01$ & $
      1.40$ & $
      0.02$ & $
    0.034$ & $
 19.5$ & 
A
 \\ 
G039.536+0.873
 & $
      4.25$ & $
      0.08$ & $
    -37.98$ & $
      0.03$ & $
      3.16$ & $
      0.07$ & $
    0.022$ & $
 48.7$ & 
C
 \\ 
G039.823+1.977
 & \nodata & \nodata & \nodata & \nodata & \nodata & \nodata & 
0.034 & 
 19.5 & 
D
 \\ 
G040.723+3.442
 & \nodata & \nodata & \nodata & \nodata & \nodata & \nodata & 
0.036 & 
 19.5 & 
D
 \\ 
\textbf{G040.955+2.473}
 & $
      2.22$ & $
      0.05$ & $
    -58.54$ & $
      0.03$ & $
      2.97$ & $
      0.08$ & $
    0.033$ & $
 19.5$ & 
A
 \\ 
\textbf{G041.304+1.997a}
 & $
      3.13$ & $
      0.02$ & $
    -57.09$ & $
      0.01$ & $
      2.16$ & $
      0.02$ & $
    0.039$ & $
 19.5$ & 
A
 \\ 
G041.304+1.997b
 & $
      2.49$ & $
      0.03$ & $
      6.87$ & $
      0.01$ & $
      1.25$ & $
      0.02$ & $
    0.039$ & $
 19.5$ & 
A
 \\ 
G041.511+1.335
 & $
      0.99$ & $
      0.02$ & $
    -46.57$ & $
      0.02$ & $
      1.71$ & $
      0.04$ & $
    0.032$ & $
 38.9$ & 
A
 \\ 
G041.522+0.827a
 & $
      1.32$ & $
      0.02$ & $
    -38.96$ & $
      0.01$ & $
      1.63$ & $
      0.03$ & $
    0.035$ & $
 19.5$ & 
B
 \\ 
G041.522+0.827b
 & $
      0.48$ & $
      0.02$ & $
    -37.10$ & $
      0.03$ & $
      1.09$ & $
      0.06$ & $
    0.035$ & $
 19.5$ & 
B
 \\ 
\textbf{G041.810+1.503a}
 & $
      0.19$ & $
      0.01$ & $
    -61.55$ & $
      0.12$ & $
      3.12$ & $
      0.31$ & $
    0.033$ & $
 19.5$ & 
B
 \\ 
G041.810+1.503b
 & $
      0.16$ & $
      0.01$ & $
    -48.87$ & $
      0.11$ & $
      3.54$ & $
      0.26$ & $
    0.033$ & $
 19.5$ & 
B
 \\ 
G042.154+1.046a
 & $
      1.86$ & $
      0.04$ & $
    -30.92$ & $
      0.02$ & $
      2.28$ & $
      0.06$ & $
    0.039$ & $
 14.6$ & 
B
 \\ 
G042.154+1.046b
 & $
      0.38$ & $
      0.04$ & $
    -45.03$ & $
      0.12$ & $
      2.31$ & $
      0.30$ & $
    0.039$ & $
 14.6$ & 
A
 \\ 
\textbf{G042.210+1.081}
 & $
      0.17$ & $
      0.01$ & $
    -57.02$ & $
      0.10$ & $
      2.74$ & $
      0.23$ & $
    0.025$ & $
 39.0$ & 
B
 \\ 
G042.224+1.205a
 & $
      0.54$ & $
      0.02$ & $
    -26.23$ & $
      0.02$ & $
      1.48$ & $
      0.06$ & $
    0.041$ & $
 19.5$ & 
C
 \\ 
G042.224+1.205b
 & $
      0.19$ & $
      0.01$ & $
    -21.14$ & $
      0.07$ & $
      1.73$ & $
      0.16$ & $
    0.041$ & $
 19.5$ & 
C
 \\ 
G042.224+1.205c
 & $
      0.18$ & $
      0.02$ & $
    -24.03$ & $
      0.06$ & $
      1.20$ & $
      0.16$ & $
    0.041$ & $
 19.5$ & 
C
 \\ 
G042.311+0.831a
 & $
      5.87$ & $
      0.04$ & $
    -30.79$ & $
      0.01$ & $
      2.68$ & $
      0.02$ & $
    0.029$ & $
 29.2$ & 
A
 \\ 
G042.311+0.831b
 & $
      0.72$ & $
      0.04$ & $
    -45.16$ & $
      0.07$ & $
      2.81$ & $
      0.17$ & $
    0.029$ & $
 29.2$ & 
A
 \\ 
G044.122+2.570
 & \nodata & \nodata & \nodata & \nodata & \nodata & \nodata & 
0.036 & 
 38.9 & 
D
 \\ 
G045.019+1.435
 & \nodata & \nodata & \nodata & \nodata & \nodata & \nodata & 
0.047 & 
  9.7 & 
D
 \\ 
G045.161+2.331
 & \nodata & \nodata & \nodata & \nodata & \nodata & \nodata & 
0.034 & 
 19.5 & 
D
 \\ 
G046.177+1.233
 & $
      1.15$ & $
      0.02$ & $
      1.92$ & $
      0.01$ & $
      1.37$ & $
      0.04$ & $
    0.027$ & $
 29.2$ & 
A
 \\ 
\textbf{G046.368+0.802a}
 & $
      4.55$ & $
      0.03$ & $
    -59.92$ & $
      0.01$ & $
      2.40$ & $
      0.02$ & $
    0.034$ & $
 19.5$ & 
A
 \\ 
G046.368+0.802b
 & $
      0.26$ & $
      0.04$ & $
    -35.60$ & $
      0.11$ & $
      1.44$ & $
      0.25$ & $
    0.034$ & $
 19.5$ & 
A
 \\ 
\textbf{G046.375+0.897a}
 & $
      4.97$ & $
      0.02$ & $
    -59.44$ & $
      0.02$ & $
      3.90$ & $
      0.04$ & $
    0.030$ & $
 38.9$ & 
C
 \\ 
\textbf{G046.375+0.897b}
 & $
      3.49$ & $
      0.04$ & $
    -63.49$ & $
      0.05$ & $
      2.81$ & $
      0.08$ & $
    0.030$ & $
 38.9$ & 
C
 \\ 
\textbf{G046.375+0.897c}
 & $
      1.08$ & $
      0.15$ & $
    -61.79$ & $
      0.04$ & $
      1.42$ & $
      0.13$ & $
    0.030$ & $
 38.9$ & 
C
 \\ 
G047.765+1.425a
 & $
      4.07$ & $
      0.04$ & $
    -35.79$ & $
      0.02$ & $
      2.27$ & $
      0.03$ & $
    0.034$ & $
 19.5$ & 
C
 \\ 
G047.765+1.425b
 & $
      2.26$ & $
      0.04$ & $
    -33.43$ & $
      0.04$ & $
      2.44$ & $
      0.06$ & $
    0.034$ & $
 19.5$ & 
C
 \\ 
G048.589+1.126
 & $
      8.70$ & $
      0.07$ & $
    -33.21$ & $
      0.01$ & $
      2.98$ & $
      0.03$ & $
    0.024$ & $
 39.0$ & 
C
 \\ 
G050.394+1.242a
 & $
      4.92$ & $
      0.05$ & $
    -53.32$ & $
      0.01$ & $
      2.06$ & $
      0.03$ & $
    0.036$ & $
 14.6$ & 
A
 \\ 
G050.394+1.242b
 & $
      0.19$ & $
      0.05$ & $
    -44.88$ & $
      0.28$ & $
      2.10$ & $
      0.67$ & $
    0.036$ & $
 14.6$ & 
B
 \\ 
G050.830+0.820
 & \nodata & \nodata & \nodata & \nodata & \nodata & \nodata & 
0.040 & 
 19.5 & 
D
 \\ 
G050.901+1.056
 & $
      1.83$ & $
      0.03$ & $
    -52.58$ & $
      0.02$ & $
      2.67$ & $
      0.05$ & $
    0.023$ & $
 39.0$ & 
A
 \\ 
G050.901+2.554
 & \nodata & \nodata & \nodata & \nodata & \nodata & \nodata & 
0.022 & 
 39.0 & 
D
 \\ 
G051.854+1.305
 & $
      0.89$ & $
      0.02$ & $
     58.37$ & $
      0.02$ & $
      2.28$ & $
      0.05$ & $
    0.041$ & $
 19.4$ & 
A
 \\ 
G052.002+1.602a
 & $
      2.70$ & $
      0.05$ & $
    -52.17$ & $
      0.02$ & $
      2.52$ & $
      0.06$ & $
    0.036$ & $
 14.6$ & 
B
 \\ 
G052.002+1.602b
 & $
      1.57$ & $
      0.05$ & $
    -56.12$ & $
      0.04$ & $
      2.44$ & $
      0.10$ & $
    0.036$ & $
 14.6$ & 
B
 \\ 
G052.021+1.629
 & \nodata & \nodata & \nodata & \nodata & \nodata & \nodata & 
0.039 & 
 19.5 & 
D
 \\ 
G052.073+2.737
 & \nodata & \nodata & \nodata & \nodata & \nodata & \nodata & 
0.022 & 
 38.9 & 
D
 \\ 
G052.706+1.526a
 & $
      2.69$ & $
      0.03$ & $
    -36.36$ & $
      0.01$ & $
      1.71$ & $
      0.02$ & $
    0.031$ & $
 19.5$ & 
A
 \\ 
G052.706+1.526b
 & $
      0.33$ & $
      0.04$ & $
      7.10$ & $
      0.06$ & $
      0.96$ & $
      0.14$ & $
    0.031$ & $
 19.5$ & 
B
 \\ 
G052.706+1.526c
 & $
      0.22$ & $
      0.04$ & $
      5.62$ & $
      0.08$ & $
      0.88$ & $
      0.21$ & $
    0.031$ & $
 19.5$ & 
B
 \\ 
G053.334+0.895
 & \nodata & \nodata & \nodata & \nodata & \nodata & \nodata & 
0.032 & 
 19.5 & 
D
 \\ 
G053.396+3.060
 & \nodata & \nodata & \nodata & \nodata & \nodata & \nodata & 
0.033 & 
 19.5 & 
D
 \\ 
G053.449+0.871
 & \nodata & \nodata & \nodata & \nodata & \nodata & \nodata & 
0.033 & 
 19.5 & 
D
 \\ 
G053.581+1.388a
 & $
      4.75$ & $
      0.02$ & $
     43.96$ & $
      0.00$ & $
      1.35$ & $
      0.01$ & $
    0.033$ & $
 19.5$ & 
A
 \\ 
G053.581+1.388b
 & $
      0.25$ & $
      0.02$ & $
     50.70$ & $
      0.06$ & $
      1.41$ & $
      0.13$ & $
    0.033$ & $
 19.5$ & 
A
 \\ 
\textbf{G054.094+1.749}
 & $
      0.52$ & $
      0.01$ & $
    -85.29$ & $
      0.02$ & $
      1.83$ & $
      0.05$ & $
    0.021$ & $
 38.9$ & 
A
 \\ 
G054.491+1.579a
 & $
      7.28$ & $
      0.06$ & $
    -38.82$ & $
      0.01$ & $
      2.37$ & $
      0.03$ & $
    0.021$ & $
 39.0$ & 
C
 \\ 
G054.491+1.579b
 & $
      1.26$ & $
      0.08$ & $
    -36.25$ & $
      0.05$ & $
      1.30$ & $
      0.11$ & $
    0.021$ & $
 39.0$ & 
C
 \\ 
G054.544+1.559a
 & $
      4.20$ & $
      0.03$ & $
    -37.29$ & $
      0.06$ & $
      3.45$ & $
      0.10$ & $
    0.023$ & $
 39.0$ & 
C
 \\ 
G054.544+1.559b
 & $
      1.36$ & $
      0.12$ & $
    -35.32$ & $
      0.02$ & $
      1.67$ & $
      0.08$ & $
    0.023$ & $
 39.0$ & 
C
 \\ 
G054.544+1.559c
 & $
      1.24$ & $
      0.13$ & $
    -39.27$ & $
      0.02$ & $
      1.61$ & $
      0.10$ & $
    0.023$ & $
 39.0$ & 
C
 \\ 
G054.544+1.559d
 & $
      0.50$ & $
      0.02$ & $
     -9.61$ & $
      0.03$ & $
      1.39$ & $
      0.06$ & $
    0.023$ & $
 39.0$ & 
A
 \\ 
G054.616+1.452a
 & $
      1.84$ & $
      0.04$ & $
    -38.27$ & $
      0.02$ & $
      1.50$ & $
      0.04$ & $
    0.033$ & $
 19.5$ & 
B
 \\ 
G054.616+1.452b
 & $
      1.58$ & $
      0.04$ & $
    -41.00$ & $
      0.02$ & $
      1.85$ & $
      0.07$ & $
    0.033$ & $
 19.5$ & 
B
 \\ 
G055.114+2.420a
 & $
      1.62$ & $
      0.04$ & $
    -66.00$ & $
      0.10$ & $
      3.38$ & $
      0.15$ & $
    0.030$ & $
 19.5$ & 
C
 \\ 
G055.114+2.420b
 & $
      1.61$ & $
      0.07$ & $
    -68.91$ & $
      0.08$ & $
      2.75$ & $
      0.11$ & $
    0.030$ & $
 19.5$ & 
C
 \\ 
G055.560+1.272a
 & $
      2.96$ & $
      0.21$ & $
     -9.86$ & $
      0.06$ & $
      1.63$ & $
      0.13$ & $
    0.033$ & $
 19.5$ & 
A
 \\ 
G055.560+1.272b
 & $
      0.38$ & $
      0.14$ & $
     44.42$ & $
      0.74$ & $
      4.16$ & $
      1.89$ & $
    0.033$ & $
 19.5$ & 
A
 \\ 
G057.107+1.457a
 & $
      4.99$ & $
      0.25$ & $
    -62.38$ & $
      0.16$ & $
      2.67$ & $
      0.25$ & $
    0.023$ & $
 38.9$ & 
C
 \\ 
G057.107+1.457b
 & $
      3.51$ & $
      0.14$ & $
     33.51$ & $
      0.03$ & $
      1.65$ & $
      0.07$ & $
    0.023$ & $
 38.9$ & 
B
 \\ 
G057.107+1.457c
 & $
      2.98$ & $
      0.65$ & $
    -60.46$ & $
      0.14$ & $
      1.81$ & $
      0.21$ & $
    0.023$ & $
 38.9$ & 
A
 \\ 
G057.107+1.457d
 & $
      1.21$ & $
      0.12$ & $
    -14.83$ & $
      0.10$ & $
      1.94$ & $
      0.23$ & $
    0.023$ & $
 38.9$ & 
C
 \\ 
G058.903+1.820
 & $
      3.14$ & $
      0.04$ & $
    -58.81$ & $
      0.01$ & $
      1.52$ & $
      0.02$ & $
    0.036$ & $
 14.6$ & 
A
 \\ 
G058.988+1.469
 & \nodata & \nodata & \nodata & \nodata & \nodata & \nodata & 
0.024 & 
 38.9 & 
D
 \\ 
G059.829+2.171
 & $
      2.68$ & $
      0.05$ & $
    -54.11$ & $
      0.02$ & $
      1.64$ & $
      0.04$ & $
    0.040$ & $
 19.5$ & 
A
 \\ 
G060.595+1.572a
 & $
      3.05$ & $
      0.12$ & $
    -47.24$ & $
      0.08$ & $
      2.71$ & $
      0.14$ & $
    0.043$ & $
  9.7$ & 
C
 \\ 
G060.595+1.572b
 & $
      1.19$ & $
      0.07$ & $
    -50.74$ & $
      0.26$ & $
      3.71$ & $
      0.52$ & $
    0.043$ & $
  9.7$ & 
C
 \\ 
\textbf{G061.085+2.502}
 & $
      1.81$ & $
      0.05$ & $
    -84.29$ & $
      0.02$ & $
      1.66$ & $
      0.05$ & $
    0.031$ & $
 19.5$ & 
A
 \\ 
G061.154+2.170a
 & $
      3.23$ & $
      0.31$ & $
    -52.47$ & $
      0.11$ & $
      2.25$ & $
      0.25$ & $
    0.035$ & $
 19.4$ & 
A
 \\ 
G061.154+2.170b
 & $
      0.44$ & $
      0.40$ & $
    -76.40$ & $
      0.62$ & $
      1.38$ & $
      1.46$ & $
    0.035$ & $
 19.4$ & 
A
 \\ 
G061.154+2.170c
 & $
      0.33$ & $
      0.28$ & $
     36.10$ & $
      1.19$ & $
      2.94$ & $
      2.81$ & $
    0.035$ & $
 19.4$ & 
A
 \\ 
\textbf{G061.180+2.447}
 & $
      4.31$ & $
      0.04$ & $
    -84.81$ & $
      0.01$ & $
      1.71$ & $
      0.02$ & $
    0.031$ & $
 19.5$ & 
A
 \\ 
G061.424+2.076a
 & $
      1.80$ & $
      0.02$ & $
    -76.84$ & $
      0.04$ & $
      3.45$ & $
      0.09$ & $
    0.030$ & $
 19.5$ & 
B
 \\ 
G061.424+2.076b
 & $
      1.54$ & $
      0.05$ & $
    -73.99$ & $
      0.03$ & $
      2.08$ & $
      0.08$ & $
    0.030$ & $
 19.5$ & 
B
 \\ 
G061.424+2.076c
 & $
      0.52$ & $
      0.02$ & $
    -70.94$ & $
      0.09$ & $
      2.97$ & $
      0.21$ & $
    0.030$ & $
 19.5$ & 
B
 \\ 
G061.587+2.074a
 & $
      0.69$ & $
      0.02$ & $
    -52.07$ & $
      0.02$ & $
      1.53$ & $
      0.05$ & $
    0.031$ & $
 19.5$ & 
A
 \\ 
G061.587+2.074b
 & $
      0.47$ & $
      0.02$ & $
     26.21$ & $
      0.03$ & $
      1.21$ & $
      0.07$ & $
    0.031$ & $
 19.5$ & 
A
 \\ 
G061.587+2.074c
 & $
      0.33$ & $
      0.02$ & $
    -49.48$ & $
      0.04$ & $
      1.11$ & $
      0.09$ & $
    0.031$ & $
 19.5$ & 
B
 \\ 
G061.587+2.074d
 & $
      0.31$ & $
      0.02$ & $
    -71.40$ & $
      0.06$ & $
      1.93$ & $
      0.13$ & $
    0.031$ & $
 19.5$ & 
B
 \\ 
G061.955+1.983
 & $
      5.90$ & $
      0.04$ & $
    -72.57$ & $
      0.01$ & $
      2.90$ & $
      0.02$ & $
    0.033$ & $
 19.5$ & 
A
 \\ 
G062.075+1.901
 & $
      0.24$ & $
      0.02$ & $
    -68.75$ & $
      0.08$ & $
      2.44$ & $
      0.19$ & $
    0.033$ & $
 19.5$ & 
A
 \\ 
G062.194+1.715a
 & $
      4.27$ & $
      0.07$ & $
    -66.29$ & $
      0.04$ & $
      2.61$ & $
      0.09$ & $
    0.031$ & $
 19.5$ & 
C
 \\ 
G062.194+1.715b
 & $
      1.21$ & $
      0.13$ & $
    -64.01$ & $
      0.08$ & $
      1.39$ & $
      0.17$ & $
    0.031$ & $
 19.5$ & 
C
 \\ 
G062.197+1.715a
 & $
      4.37$ & $
      0.05$ & $
    -66.50$ & $
      0.02$ & $
      2.82$ & $
      0.05$ & $
    0.023$ & $
 38.9$ & 
C
 \\ 
G062.197+1.715b
 & $
      0.95$ & $
      0.07$ & $
    -63.95$ & $
      0.06$ & $
      1.44$ & $
      0.14$ & $
    0.023$ & $
 38.9$ & 
C
 \\ 
G062.325+3.020a
 & $
      2.95$ & $
      0.06$ & $
    -70.49$ & $
      0.01$ & $
      2.96$ & $
      0.03$ & $
    0.032$ & $
 19.4$ & 
C
 \\ 
G062.325+3.020b
 & $
      2.50$ & $
      0.06$ & $
    -71.03$ & $
      0.01$ & $
      1.20$ & $
      0.02$ & $
    0.032$ & $
 19.4$ & 
C
 \\ 
G062.325+3.020c
 & $
      0.95$ & $
      0.02$ & $
      7.79$ & $
      0.02$ & $
      1.62$ & $
      0.04$ & $
    0.032$ & $
 19.4$ & 
A
 \\ 
\textbf{G062.578+2.387a}
 & $
      9.56$ & $
      0.03$ & $
    -69.20$ & $
      0.01$ & $
      3.76$ & $
      0.03$ & $
    0.023$ & $
 38.9$ & 
C
 \\ 
G062.578+2.387b
 & $
      3.21$ & $
      0.05$ & $
    -65.33$ & $
      0.04$ & $
      2.99$ & $
      0.07$ & $
    0.023$ & $
 38.9$ & 
A
 \\ 
G062.578+2.387c
 & $
      0.57$ & $
      0.06$ & $
    -89.09$ & $
      0.06$ & $
      1.29$ & $
      0.16$ & $
    0.023$ & $
 38.9$ & 
C
 \\ 
G062.819+3.144
 & $
      5.03$ & $
      0.04$ & $
    -69.92$ & $
      0.01$ & $
      1.40$ & $
      0.01$ & $
    0.034$ & $
 19.5$ & 
A
 \\ 
G064.151+1.282a
 & $
      5.30$ & $
      0.08$ & $
    -54.42$ & $
      0.04$ & $
      2.95$ & $
      0.05$ & $
    0.023$ & $
 38.9$ & 
C
 \\ 
G064.151+1.282b
 & $
      1.83$ & $
      0.15$ & $
    -56.54$ & $
      0.07$ & $
      2.33$ & $
      0.08$ & $
    0.023$ & $
 38.9$ & 
C
 \\ 
G064.151+1.282c
 & $
      0.48$ & $
      0.02$ & $
    -24.43$ & $
      0.05$ & $
      2.33$ & $
      0.11$ & $
    0.023$ & $
 38.9$ & 
A
 \\ 
G064.862+3.120a
 & $
     11.50$ & $
      0.07$ & $
     -1.79$ & $
      0.01$ & $
      1.99$ & $
      0.01$ & $
    0.031$ & $
 19.5$ & 
A
 \\ 
G064.862+3.120b
 & $
      0.26$ & $
      0.07$ & $
     26.67$ & $
      0.28$ & $
      2.22$ & $
      0.65$ & $
    0.031$ & $
 19.5$ & 
A
 \\ 
G066.608+2.061a
 & $
      5.16$ & $
      0.39$ & $
    -67.69$ & $
      0.09$ & $
      2.55$ & $
      0.08$ & $
    0.031$ & $
 19.5$ & 
C
 \\ 
G066.608+2.061b
 & $
      2.54$ & $
      0.30$ & $
    -69.76$ & $
      0.22$ & $
      2.77$ & $
      0.24$ & $
    0.031$ & $
 19.5$ & 
C
 \\ 
G066.608+2.061c
 & $
      0.46$ & $
      0.04$ & $
    -50.57$ & $
      0.08$ & $
      1.97$ & $
      0.19$ & $
    0.031$ & $
 19.5$ & 
A
 \\ 
G067.138+1.966
 & $
      6.10$ & $
      0.11$ & $
    -71.31$ & $
      0.02$ & $
      1.80$ & $
      0.04$ & $
    0.024$ & $
 38.9$ & 
C
 \\ 
 \enddata
 \tablecomments{Names in bold face have \(\vlsr\) within OSC velocity
   range defined by $V_{\rm LSR} = -1.6\,\ell \pm 15\kms$.}
\end{deluxetable*}

\subsection{CO Distribution throughout the Milky Way}

How are the molecular clouds with detected CO emission distributed in
the Galaxy?  Our targets were selected from the {\it WISE}
\hii\ region catalog and therefore their distribution should reflect
the Galactic \hii\ region or high-mass star formation distribution.
We expect that in many cases the CO emission is associated with the
\hii\ region along the line-of-sight, but this is not always true
\citep{russeil2004,anderson2009}.

The most direct view is given by the observed, model independent,
parameters: $\ell, b, v$.  Figure~\ref{fig:lv_diagram} shows the
($\ell, b$) (top) and the ($\ell, v$) diagrams (bottom) for the OSC
region of the Galaxy.  Similar plots are found in \citet{dame2011} and
\citet{armentrout2017}.  Integrated \hi\ emission is shown together
with our \co\ detections (red circles).  The molecular clouds trace
the warp in the disk as defined by \hi\ emission but are concentrated
at the lower latitude end of the \hi\ envelope.  The bottom panel of
Figure~\ref{fig:lv_diagram} shows the molecular clouds located in the
OSC as red circles between the dashed lines. Several of these are
clearly associated with \hii\ regions indicated by the gray stars.

Since the majority of our targets have multiple emission components,
we cannot identify which CO component might be associated with the
\hii\ region.  Nevertheless, most of the molecular clouds in our
sample have negative velocities and are thus located beyond the Solar
orbit. In fact, Figure~\ref{fig:lv_diagram} shows that the majority of
the negative velocity components CO seem to be located in the Outer
Arm and not the more-distant OSC.

\startlongtable
\begin{deluxetable*}{lrrrrrrrrr}
\tabletypesize{\scriptsize}
\tablecaption{\cor Emission Line Parameters\label{tab:co13}}
\tablehead{
\colhead{Name} & 
\colhead{$T_L$} & 
\colhead{$\sigma_{T_L}$} & 
\colhead{$V_{\rm LSR}$} & 
\colhead{$\sigma_{V_{\rm LSR}}$} & 
\colhead{$\Delta V$} & 
\colhead{$\sigma_{\Delta V}$} & 
\colhead{RMS} & 
\colhead{$t_{\rm intg}$} & 
\colhead{QF}\\
\colhead{} & 
\colhead{(K)} & 
\colhead{(K)} & 
\colhead{(\kms)} & 
\colhead{(\kms)} & 
\colhead{(\kms)} & 
\colhead{(\kms)} & 
\colhead{(K)} & 
\colhead{(min.)} & 
\colhead{}
}
\startdata
\textbf{G033.008+1.151}
 & $
      0.36$ & $
      0.01$ & $
    -47.04$ & $
      0.03$ & $
      2.93$ & $
      0.06$ & $
    0.010$ & $
 77.9$ & 
A
 \\ 
G038.627+0.813
 & \nodata & \nodata & \nodata & \nodata & \nodata & \nodata & 
0.011 & 
 68.2 & 
D
 \\ 
\textbf{G039.185$-$1.421a}
 & $
      0.33$ & $
      0.01$ & $
    -55.30$ & $
      0.03$ & $
      2.87$ & $
      0.08$ & $
    0.012$ & $
 58.4$ & 
A
 \\ 
\textbf{G040.955+2.473}
 & $
      0.19$ & $
      0.01$ & $
    -58.60$ & $
      0.05$ & $
      2.62$ & $
      0.12$ & $
    0.011$ & $
 63.3$ & 
A
 \\ 
\textbf{G041.304+1.997a}
 & $
      0.30$ & $
      0.01$ & $
    -57.11$ & $
      0.02$ & $
      1.77$ & $
      0.04$ & $
    0.011$ & $
 58.4$ & 
A
 \\ 
G041.304+1.997b
 & $
      0.26$ & $
      0.01$ & $
      6.95$ & $
      0.04$ & $
      1.14$ & $
      0.08$ & $
    0.011$ & $
 58.4$ & 
A
 \\ 
G041.511+1.335
 & $
      0.15$ & $
      0.00$ & $
    -46.67$ & $
      0.02$ & $
      1.23$ & $
      0.05$ & $
    0.011$ & $
 58.4$ & 
A
 \\ 
\textbf{G046.368+0.802a}
 & $
      0.63$ & $
      0.01$ & $
    -60.13$ & $
      0.03$ & $
      2.46$ & $
      0.06$ & $
    0.021$ & $
 19.5$ & 
A
 \\ 
\textbf{G046.375+0.897a}
 & $
      0.64$ & $
      0.01$ & $
    -59.42$ & $
      0.07$ & $
      3.73$ & $
      0.14$ & $
    0.011$ & $
 68.1$ & 
C
 \\ 
\textbf{G046.375+0.897b}
 & $
      0.36$ & $
      0.01$ & $
    -63.51$ & $
      0.10$ & $
      3.09$ & $
      0.22$ & $
    0.011$ & $
 68.1$ & 
C
 \\ 
G050.394+1.242a
 & $
      0.61$ & $
      0.01$ & $
    -53.55$ & $
      0.01$ & $
      1.81$ & $
      0.02$ & $
    0.020$ & $
 19.5$ & 
A
 \\ 
G050.901+1.056a
 & $
      0.26$ & $
      0.01$ & $
    -51.89$ & $
      0.04$ & $
      1.32$ & $
      0.07$ & $
    0.011$ & $
 58.4$ & 
C
 \\ 
G050.901+1.056b
 & $
      0.12$ & $
      0.01$ & $
    -53.37$ & $
      0.13$ & $
      1.71$ & $
      0.23$ & $
    0.011$ & $
 58.4$ & 
C
 \\ 
G052.002+1.602a
 & $
      0.34$ & $
      0.01$ & $
    -52.56$ & $
      0.03$ & $
      2.17$ & $
      0.08$ & $
    0.011$ & $
 58.4$ & 
B
 \\ 
G052.002+1.602b
 & $
      0.17$ & $
      0.01$ & $
    -55.81$ & $
      0.07$ & $
      2.18$ & $
      0.21$ & $
    0.011$ & $
 58.4$ & 
B
 \\ 
G054.094+1.749
 & \nodata & \nodata & \nodata & \nodata & \nodata & \nodata & 
0.009 & 
 77.9 & 
D
 \\ 
G055.114+2.420b
 & $
      0.12$ & $
      0.01$ & $
    -68.67$ & $
      0.04$ & $
      1.34$ & $
      0.11$ & $
    0.011$ & $
 58.4$ & 
C
 \\ 
G055.114+2.420a
 & $
      0.11$ & $
      0.01$ & $
    -67.62$ & $
      0.11$ & $
      4.64$ & $
      0.19$ & $
    0.011$ & $
 58.4$ & 
C
 \\ 
G057.107+1.457a
 & $
      1.35$ & $
      0.02$ & $
    -61.67$ & $
      0.02$ & $
      2.58$ & $
      0.04$ & $
    0.019$ & $
 19.5$ & 
A
 \\ 
G058.903+1.820
 & $
      0.36$ & $
      0.02$ & $
    -58.78$ & $
      0.03$ & $
      1.26$ & $
      0.07$ & $
    0.012$ & $
 48.7$ & 
A
 \\ 
G059.829+2.171
 & $
      0.75$ & $
      0.01$ & $
    -54.17$ & $
      0.01$ & $
      1.26$ & $
      0.02$ & $
    0.015$ & $
 24.3$ & 
A
 \\ 
G060.595+1.572a
 & $
      0.28$ & $
      0.01$ & $
    -47.45$ & $
      0.03$ & $
      1.50$ & $
      0.08$ & $
    0.012$ & $
 48.7$ & 
C
 \\ 
G060.595+1.572b
 & $
      0.14$ & $
      0.01$ & $
    -48.61$ & $
      0.15$ & $
      6.56$ & $
      0.33$ & $
    0.012$ & $
 48.7$ & 
C
 \\ 
G061.154+2.170a
 & $
      0.82$ & $
      0.01$ & $
    -52.44$ & $
      0.01$ & $
      1.80$ & $
      0.03$ & $
    0.016$ & $
 19.6$ & 
A
 \\ 
G061.154+2.170d
 & $
      0.53$ & $
      0.01$ & $
     11.67$ & $
      0.05$ & $
      2.39$ & $
      0.11$ & $
    0.016$ & $
 19.6$ & 
A
 \\ 
G061.154+2.170e
 & $
      0.22$ & $
      0.02$ & $
      4.58$ & $
      0.04$ & $
      1.06$ & $
      0.09$ & $
    0.016$ & $
 19.6$ & 
A
 \\ 
\textbf{G061.180+2.447}
 & $
      0.60$ & $
      0.01$ & $
    -84.74$ & $
      0.01$ & $
      1.35$ & $
      0.02$ & $
    0.017$ & $
 19.5$ & 
A
 \\ 
G061.424+2.076a
 & $
      0.40$ & $
      0.05$ & $
    -76.59$ & $
      0.21$ & $
      2.57$ & $
      0.26$ & $
    0.017$ & $
 19.4$ & 
C
 \\ 
G061.424+2.076b
 & $
      0.24$ & $
      0.04$ & $
    -74.24$ & $
      0.39$ & $
      2.76$ & $
      0.49$ & $
    0.017$ & $
 19.4$ & 
C
 \\ 
G061.955+1.983
 & $
      1.23$ & $
      0.01$ & $
    -72.64$ & $
      0.01$ & $
      2.34$ & $
      0.03$ & $
    0.017$ & $
 19.5$ & 
A
 \\ 
G062.075+1.901
 & \nodata & \nodata & \nodata & \nodata & \nodata & \nodata & 
0.010 & 
 58.4 & 
D
 \\ 
G062.194+1.715a1
 & $
      0.44$ & $
      0.03$ & $
    -66.09$ & $
      0.03$ & $
      1.26$ & $
      0.07$ & $
    0.010$ & $
 58.3$ & 
C
 \\ 
G062.194+1.715a2
 & $
      0.24$ & $
      0.01$ & $
    -67.49$ & $
      0.08$ & $
      1.80$ & $
      0.13$ & $
    0.010$ & $
 58.3$ & 
C
 \\ 
G062.194+1.715b
 & $
      0.21$ & $
      0.01$ & $
    -64.66$ & $
      0.09$ & $
      1.86$ & $
      0.13$ & $
    0.010$ & $
 58.3$ & 
C
 \\ 
G062.197+1.715a
 & $
      0.80$ & $
      0.02$ & $
    -66.09$ & $
      0.04$ & $
      2.78$ & $
      0.09$ & $
    0.017$ & $
 19.4$ & 
A
 \\ 
G062.325+3.020b
 & $
      1.15$ & $
      0.04$ & $
    -70.78$ & $
      0.03$ & $
      1.23$ & $
      0.04$ & $
    0.018$ & $
 19.5$ & 
C
 \\ 
G062.325+3.020a
 & $
      0.57$ & $
      0.03$ & $
    -69.61$ & $
      0.07$ & $
      1.37$ & $
      0.09$ & $
    0.018$ & $
 19.5$ & 
C
 \\ 
G062.578+2.387a
 & $
      1.88$ & $
      0.01$ & $
    -69.28$ & $
      0.01$ & $
      3.04$ & $
      0.03$ & $
    0.018$ & $
 19.5$ & 
C
 \\ 
G062.578+2.387b
 & $
      0.46$ & $
      0.01$ & $
    -65.55$ & $
      0.04$ & $
      2.52$ & $
      0.10$ & $
    0.018$ & $
 19.5$ & 
C
 \\ 
G062.819+3.144
 & $
      0.68$ & $
      0.01$ & $
    -69.84$ & $
      0.01$ & $
      1.25$ & $
      0.02$ & $
    0.017$ & $
 19.4$ & 
A
 \\ 
G064.151+1.282a
 & $
      0.64$ & $
      0.01$ & $
    -54.88$ & $
      0.02$ & $
      2.91$ & $
      0.05$ & $
    0.016$ & $
 19.5$ & 
A
 \\ 
G066.608+2.061ab
 & $
      1.60$ & $
      0.04$ & $
    -68.21$ & $
      0.03$ & $
      2.42$ & $
      0.08$ & $
    0.017$ & $
 19.5$ & 
A
 \\ 
G067.138+1.966
 & $
      1.26$ & $
      0.01$ & $
    -71.47$ & $
      0.01$ & $
      1.37$ & $
      0.02$ & $
    0.017$ & $
 19.5$ & 
A
 \\ 
\enddata
 \tablecomments{Names in bold face have \(\vlsr\) within OSC velocity
   range defined by $V_{\rm LSR} = -1.6\,\ell \pm 15\kms$.}
\end{deluxetable*}

\startlongtable
\begin{deluxetable*}{lrrcrrr}
\tabletypesize{\scriptsize}
\tablecaption{Molecular Cloud Properties \label{tab:prop}}
\tablewidth{0pt}
\tablehead{
\colhead{Name} & \colhead{$\vlsr$} &  \colhead{$R_{\rm gal}$} & 
\colhead{$d_\odot$} & \colhead{$T_{\rm L}$($^{12}$CO)} & \colhead{$T_{\rm L}$($^{13}$CO)} & \colhead{$N({\rm H}_{2})$} \\ 
\colhead{} & \colhead{(km$\,$s$^{-1}$)} & \colhead{(kpc)} & 
\colhead{(kpc)} & \colhead{(K)} & \colhead{(K)} & \colhead{($10^{20}\,$cm$^{-2}$)} } 
\startdata 
G033.008+1.151 & $-$46.84 & 14.44 & 20.81 & 3.75 & 0.36 & 5.19 \\
G039.185$-$1.421a & $-$55.67 & 14.69 & 20.26 & 1.52 & 0.33 & 4.66 \\
G040.955+2.473 & $-$58.54 & 14.83 & 20.16 & 2.22 & 0.19 & 2.45 \\
G041.304+1.997a & $-$57.09 & 14.50 & 19.76 & 3.13 & 0.30 & 2.61 \\
G041.304+1.997b & 6.87 & 8.22 & 0.38/12.39 & 2.49 & 0.26 & 1.46 \\
G041.511+1.335 & $-$46.57 & 12.87 & 17.94 & 0.99 & 0.15 & 0.91 \\
G046.368+0.802a & $-$59.92 & 14.09 & 18.55 & 4.55 & 0.63 & 7.63 \\
G046.375+0.897a & $-$59.44 & 14.02 & 18.46 & 4.97 & 0.64 & 11.75 \\
G046.375+0.897b & $-$63.49 & 14.64 & 19.14 & 3.49 & 0.36 & 5.47 \\
G050.394+1.242a & $-$53.32 & 12.78 & 16.39 & 4.92 & 0.61 & 5.43 \\
G050.901+1.056a & $-$52.58 & 12.65 & 16.16 & 1.83 & 0.26 & 1.69 \\
G050.901+1.056b & $-$52.58 & 12.65 & 16.16 & 1.83 & 0.12 & 1.01 \\
G052.002+1.602a & $-$52.17 & 12.52 & 15.81 & 2.70 & 0.34 & 3.63 \\
G052.002+1.602b & $-$56.12 & 12.96 & 16.32 & 1.57 & 0.17 & 1.82 \\
G055.114+2.420a & $-$66.00 & 13.86 & 16.84 & 1.62 & 0.11 & 2.51 \\
G055.114+2.420b & $-$68.91 & 14.23 & 17.27 & 1.61 & 0.12 & 0.79 \\
G057.107+1.457a & $-$62.38 & 13.24 & 15.77 & 4.99 & 1.35 & 17.14 \\
G058.903+1.820 & $-$58.81 & 12.73 & 14.84 & 3.14 & 0.36 & 2.23 \\
G059.829+2.171 & $-$54.11 & 12.22 & 14.04 & 2.68 & 0.75 & 4.65 \\
G060.595+1.572a & $-$47.24 & 11.58 & 13.08 & 3.05 & 0.28 & 2.07 \\
G060.595+1.572b & $-$50.74 & 11.88 & 13.46 & 1.19 & 0.14 & 4.52 \\
G061.154+2.170a & $-$52.47 & 12.01 & 13.52 & 3.23 & 0.82 & 7.26 \\
G061.180+2.447 & $-$84.81 & 15.77 & 18.00 & 4.31 & 0.60 & 3.99 \\
G061.424+2.076a & $-$76.84 & 14.62 & 16.64 & 1.80 & 0.40 & 5.06 \\
G061.424+2.076b & $-$73.99 & 14.25 & 16.21 & 1.54 & 0.24 & 3.26 \\
G061.955+1.983 & $-$72.57 & 14.04 & 15.86 & 5.90 & 1.23 & 14.16 \\
G062.194+1.715a1 & $-$66.29 & 13.30 & 14.94 & 4.27 & 0.44 & 2.73 \\
G062.194+1.715a2 & $-$66.29 & 13.30 & 14.94 & 4.27 & 0.24 & 2.13 \\
G062.194+1.715b & $-$64.01 & 13.06 & 14.64 & 1.21 & 0.21 & 1.92 \\
G062.197+1.715a & $-$66.50 & 13.33 & 14.97 & 4.37 & 0.80 & 10.94 \\
G062.325+3.020a & $-$70.49 & 13.77 & 15.48 & 2.95 & 0.57 & 3.84 \\
G062.325+3.020b & $-$71.03 & 13.83 & 15.55 & 2.50 & 1.15 & 6.96 \\
G062.578+2.387a & $-$69.20 & 13.60 & 15.23 & 9.56 & 1.88 & 28.12 \\
G062.578+2.387b & $-$65.33 & 13.18 & 14.72 & 3.21 & 0.46 & 5.70 \\
G062.819+3.144 & $-$69.92 & 13.67 & 15.27 & 5.03 & 0.68 & 4.18 \\
G064.151+1.282a & $-$54.42 & 12.05 & 13.02 & 5.30 & 0.64 & 9.16 \\
G067.138+1.966 & $-$71.31 & 13.54 & 14.35 & 6.10 & 1.26 & 8.49 \\
\enddata 
\tablecomments{Sources within the Solar circle have two values for the Distance
 (d$_\odot$) since we did not resolve the kinematic distance ambiguity.}
\end{deluxetable*}

Face-on maps of the Milky Way disk are difficult to make since
distances to the sources are required.  Parallax distances are the
most accurate but can only be measured for a relatively small sample
\citep[e.g., see][]{reid2014}.  Kinematic distances are far easier to
derive since they only require a spectral line velocity and a model of
Galactic rotation.  The main limitations are that the rotation curves
suffer from non-circular (streaming) motions, and that within the
Solar orbit there is a kinematic distance ambiguity
\citep[e.g.,][]{anderson2012}.  Studies of the OSC arm avoid the
latter problem because the OSC is located beyond the Solar orbit and
so does not suffer from the distance ambiguity issue.

\begin{figure}[!htb]
\centering
\includegraphics[width=\linewidth]{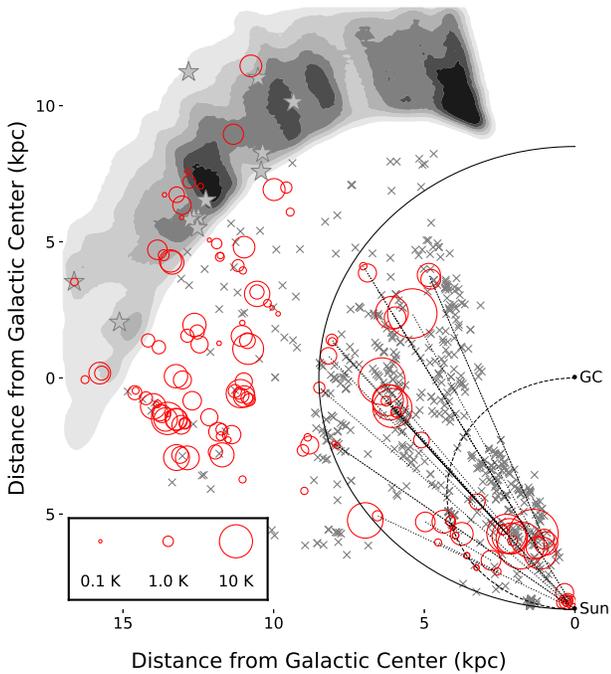}
\caption{Face-on map of \hii\ regions, CO emission components, and
  \hi\ emission in the first Galactic quadrant.  Plotted are
  \hii\ regions from the {\it WISE} \hii\ region catalog (gray
  crosses), \hii\ regions consistent with the location of the OSC
  (gray stars), and CO emission detected here in the direction of {\it
    WISE} \hii\ regions (red circles).  The red circle sizes are
  scaled by the intensity of the \co\ peak line intensity.  The Solar
  circle and tangent points are indicated by the solid and dashed
  curves, respectively.  The dotted lines connect the near and far
  kinematic distances for sources within the Solar orbit since we did
  not resolve the kinematic distance ambiguity.  The gray scale shows
  OSC \hi\ emission from the LAB survey. Here the OSC is defined in
  latitude and velocity as: $V_{\rm LSR} = -1.6\,{\rm km}\,{\rm
    s}^{-1}\,{\rm deg}^{-1} \times \ell$ and $b = $0\degper 75 + 0.75
  $\times \ell$.  This is the same \hi\ emission shown between the
  dashed lines in Figure~\ref{fig:lv_diagram} (bottom panel), but
  transformed using the Brand rotation curve
  \citep{brand1993}. (Reproduced from \citet{armentrout2017}.)}
\label{fig:faceon_diagram}
\end{figure}

We use the \citet{brand1993} rotation curve to determine kinematic
distances to \hi\ and CO emission as well as \hii\ regions.  The
resulting face-on map for the first quadrant OSC zone studied here is
shown in Figure~\ref{fig:faceon_diagram} where the gray crosses are
taken from the {\it WISE} \hii\ region catalog and the red circles are
CO emission from this study. We also include previously detected OSC
\hii\ regions (gray stars) and \hi\ data (gray contours). The
\hii\ regions and molecular clouds associated with the OSC span a wide
range of longitudes in the first Galactic quadrant.  Their
distribution is well-matched to the OSC extent delineated by
\hi\ emission.

\begin{figure}[!htb]
\centering
\includegraphics[width=\linewidth]{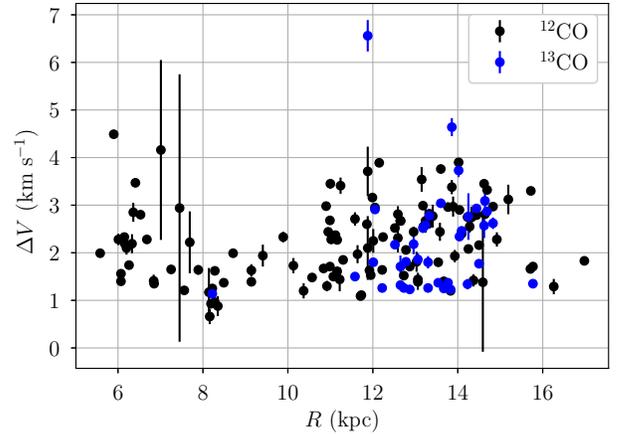}
\caption{\co\ (black) and \cor\ (blue) emission line FWHM line width,
  \(\Delta V\), as a function of Galactocentric radius, \(R\). The
  black circles and black squares are the \co\ and \cor\ FWHM line
  widths, respectively. The error bars are the \(1\sigma\)
  uncertainties in the measured FWHM line widths.}
\label{fig:fwhm_vs_R}
\end{figure}

\begin{figure}[!htb]
\centering
\includegraphics[width=\linewidth]{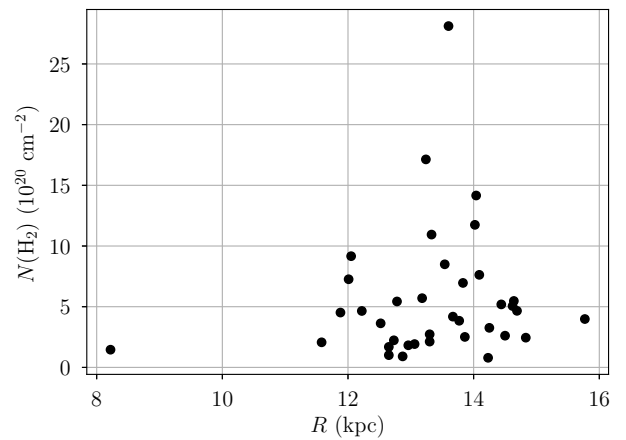}
\caption{Molecular column density, \(N({\rm H}_2)\), as a function of
  Galactocentric radius, \(R\), for molecular clouds with
  \cor\ detections.}
\label{fig:NH2_vs_R}
\end{figure}

\subsection{Comparison of OSC and non-OSC CO Clouds}

We search for differences between the emission line properties of
molecular clouds in the OSC and clouds not in the OSC.
Figure~\ref{fig:fwhm_vs_R} shows the \co\ and \cor\ FWHM line widths
as a function of Galactocentric radius, \(R\). The ratio of the mean
\co\ FWHM line width of OSC clouds to non--OSC clouds is \(1.13\) with
a standard deviation of \(0.52\). For \cor\, the ratio is \(1.23\)
with a standard deviation of \(0.74\). This evidence suggests
molecular cloud velocity dispersions are fairly uniform for all
molecular clouds in our sample. We note, however, that our
observations are a single pointing in the direction of the cloud
rather than a full map. Our derived FWHM line widths are therefore
likely lower limits of the true molecular cloud FWHM line widths,
especially if the clouds are much larger than our beam
(\({\sim}1\)\arcmin\ ).

The \cor\ emission lines should be optically thin and may be used to
derive the total molecular column densities of the molecular clouds.
We derive the total H\(_2\) column density of each of our
\cor\ detections using Equation (2) from \citet{simon2001}. This
equation assumes a \co/\cor\ abundance ratio of 45, an excitation
temperature of \(10\,\K\), and an X-factor (\co/H$_{2}$) of
\nexpo{8}{-5}.  We assume the main beam efficiency is near unity and
use the \cor\ emission line properties from Table~\ref{tab:co13}.  The
molecular cloud column densities are compiled in Table~\ref{tab:prop}
where we list the \co\ component name, the LSR velocity, the
Galactocentric radius, the distance from the Sun, the \co\ and
\cor\ line intensities, and the molecular column density.  Sources
within the Solar circle have two values for the distance since we did
not resolve the kinematic distance ambiguity.

The total molecular column density of these clouds also does not have
any trend with \(R\). Figure~\ref{fig:NH2_vs_R} shows the derived
H\(_2\) column densities as a function of \(R\) for each of our clouds
with \cor\ detections. The ratio of the mean molecular column
densities of OSC clouds to non--OSC clouds is \(0.97\) with a standard
deviation of \(1.11\). This evidence again suggests no difference
between OSC and non--OSC molecular clouds within our sample with the
caveat that our pointed observations likely underestimate the true
molecular cloud FWHM line widths, and therefore these derived column
densities may represent lower limits.

Since we expect molecular clouds to be extended in CO emission and
have not mapped these regions, we cannot accurately derive the total
molecular mass of these complexes.  The eight sources from the MWISP
survey in common with our survey have molecular masses and radii that
range from ${\sim}200-7000$\msun\ and ${\sim}3-13$\pc, respectively.
Since the molecular masses were calculated using \co\ intensities that
are likely optically thick, these should be interpreted as lower
limits.

\section{Conclusion}

The Outer Scutum-Centaurus (OSC) spiral arm is the most distant site
of massive star formation in the Milky Way. The OSC contains at least
12 known \hii\ regions ionized by high-mass stars with spectral types
as early as O4.  Molecular clouds have been detected within the OSC
but most have not been associated with \hii\ regions.  Here we use the
Arizona Radio Observatory (ARO) 12\m\ telescope to observe CO emission
from 78 {\it WISE} \hii\ region candidates located within the Galactic
longitude-latitude ($\ell, b$) locus of the OSC spiral arm in the
first quadrant ($20^{\circ} < \ell < 90^{\circ}$).  We detect 117
\co\ spectral line components in 62 of 78 targets, and 40
\cor\ components in 27 of 30 targets.  About 2/3 of the molecular
clouds reside beyond the Solar orbit and are associated with the Outer
Arm. We discovered 17 \co\ emission lines and 8 \cor\ emission lines
consistent with the ($\ell, v$) OSC locus. These OSC molecular clouds
have the same physical properties (FWHM line widths and molecular
column densities) as non--OSC clouds within our sample.

\acknowledgments

We thank T. M. Dame and the anonymous referee for their useful
comments and suggestions which improved the quality of this
paper. T.V.W. is supported by the NSF through the Grote Reber
Fellowship Program administered by Associated Universities,
Inc./National Radio Astronomy Observatory, the D.N. Batten Foundation
Fellowship from the Jefferson Scholars Foundation, the Mars Foundation
Fellowship from the Achievement Rewards for College Scientists
Foundation, and the Virginia Space Grant Consortium.  L.D.A. is
supported by NSF grant AST1516021. The Kitt Peak 12 Meter is operated
by the Arizona Radio Observatory (ARO), Steward Observatory,
University of Arizona. \nraoblurb

\facilities{ARO$\,12\,$m}

\software{TMBIDL \citep{bania2016}}

\bibliography{co}

\end{document}